\documentstyle[12pt,epsf]{article}

\newcommand {\be}       {\begin{equation}}
\newcommand {\ee}    {\end{equation}}
\newcommand {\bea}      {\begin{eqnarray}}
\newcommand {\eea}  {\end{eqnarray}}
\newcommand {\tf}[1]    {\mbox{\em #1}}              

\newcommand {\bib}[1]   {\bibitem{#1}}

\newcommand {\tr}       {\mbox{tr }}

\newcommand {\e}[1]     {\mbox{e}^{#1}}

\newcommand {\sqbr}[1]  {\left[ #1 \right] }
\begin{document}
\begin{center} 
\section*{The Coupled Cluster Method
in Hamiltonian Lattice Field Theory}

 D.Sch\"{u}tte\\
Institut f\"{u}r Theoretische Kernphysik der Universit\"{a}t Bonn \\
         Nu{\ss}allee 14 -- 16, D-53115 Bonn, Germany \\
and\\
Zheng Weihong and C.J. Hamer\\
School of Physics, The University of New South Wales\\
Sydney, NSW 2052, Australia,\\
\end{center}

\begin{abstract}
The coupled cluster
or  exp S form of the
eigenvalue problem for 
lattice Hamiltonian QCD (without quarks) is investigated.

A new construction prescription 
is given for the calculation
of the relevant coupled cluster matrix elements
with respect to an orthogonal
and independent loop space basis.
The method avoids 
 the explicit introduction of gauge group
coupling coefficients by mapping the 
eigenvalue problem onto a suitable set of
character functions, which allows a simplified procedure.

Using appropriate group theoretical methods,
we show that it is possible to set
up the eigenvalue problem for eigenstates having
arbitrary lattice momentum and lattice angular 
momentum.

\end{abstract}
\newpage

\section{Introduction and Overview}

The investigation of the eigenvalue problem for the 
lattice QCD Hamiltonian
is considered to be an alternative to standard 
Lagrangian lattice Monte Carlo QCD,
possibly giving  new
 insight into the structure of such non-abelian gauge theories.

For pure SU(3) Yang-Mills theory\cite{KoSu75} (without fermions)
in particular,
many attempts have been made to attack the
corresponding Kogut-Susskind Hamiltonian problem:
for instance, there exist 
the strong coupling expansion\cite{Hamer},
the exp(-tH) method\cite{Horn} or 
variational techniques\cite{Chin}.
Up to now, none of these approaches could
obtain results for excited states (e.g. glueball masses) 
comparable
in control and accuracy to those
within the Euclidean Monte Carlo method
(there has been, however, some  progress for
ground states using the Greens function Monte Carlo
method\cite{Weihong}).

This also holds for 
the coupled cluster (exp S) method
which  attracted
special attention in recent years\cite{Llew,Guo,Luo,Mak}.
(Some   encouraging results 
within this framework were obtained
recently\cite{Guo95}.)  Here the basic idea
is to incorporate manifestly the correct volume dependencies
of observables by writing the ground state in the form
$\psi_0 = \e{S}$ and putting
$\psi = F \psi_0$ for excited states.
The "Schroedinger" equation for the functions $S$ and $F$
can be formulated rigorously\cite{Llew} and it is
tempting to define approximations by a suitable
truncation of a loop space expansion of these
quantities\cite{Llew,Guo,Guo95}.

It is the purpose of this paper to further 
elucidate the structure
of this  coupled cluster method
with the hope that 
the resulting insights  may lead
to improved calculations of the QCD
spectrum.

We will concentrate our considerations 
on the treatment of the Kogut-Susskind
Hamiltonian as the lattice regularization 
of an $SU(n)$ Yang Mills theory.
A discussion of the full QCD
and its treatment within 
a quenched approximation
is possible, but this will be
deferred to a future publication.

We now give an outline of our paper
which summarizes at the same time
our methods and our results.

Our basic tools will be group theoretical methods
which will be introduced in section 2.
The group of the link variables, the local
lattice gauge group and the lattice Euclidean
group will play a role.

As discussed in section 3,
projection operators on representations of the lattice 
Euclidean group  with given lattice momentum
and lattice angular momentum allow one to introduce the
notion of an intrinsic wavefunction 
related to the ground state function $S$ and
to the ``excitation operator'' $F$.

This structure
has been used in Refs.\cite{Llew,Guo,Guo95} for
the trivial representation; here we provide
a systematic  framework for
 general representations of the lattice Euclidean group.

The solution of the eigenvalue problem for the
Kogut-Susskind Hamiltonian is then
reduced to the determination
of the intrinsic eigenfunctions.

For this purpose, a basis of suitable wavefunctions is
needed which  may be used for an expansion
and which allows a computation of the relevant
coupled cluster matrix elements.
Within the Kogut-Susskind theory these have
to be functions of the link variables 
which are invariant under the action of the
local lattice gauge group.
 
The problem of setting up such a basis
in an effective way is addressed
in section. 4.
There exist two strategies for the construction
of such basis systems:

1) Choose first a basis for the functions
of the individual link variables  given by 
the standard D-functions.
General polynomials of these functions with
different link variables, combined with
suitable $SU(n)$ coupling coefficients,  
form then the desired basis for the
intrinsic hadron (or vacuum) wave functions.
We call this set of functions the {\em D-loop basis}.

Details of this construction have been
worked out in Ref.\cite{Irv1}. An application is
the "exact linked
cluster expansion"
discussed in Ref.\cite{Irving}.

This method is limited by the necessity 
to handle an increasing number
of $SU(n)$ coupling coefficients.

A clear merit of the procedure is that
it provides an independent,
 orthogonal and (in the limit
of increasing polynomial degree)
complete basis of physical states.

2) An alternative  system 
of physical states is provided by the set of
{\em character functions} corresponding
to an expansion in terms
of suitable group characters.
This approach was used in the recent
calculations within the coupled cluster method\cite{Llew,Guo}.
The obvious advantage here is that each term
is manifestly locally gauge invariant,
and no coupling with $SU(n)$ Clebsch Gordon
coefficients is needed.
The problem, however, is that the emerging
system of wavefunctions is in general non-orthogonal
and {\em overcomplete}.

In Refs.\cite{Llew,Guo,Luo} the disease of having linear
dependencies was cured with the help of
a special form of
the Cayley-Hamilton relation for $SU(n)$ matrices.
This method, however,  does not appear to be very
systematic, and only calculations with wavefunctions
generated from up to fourth order 
plaquette polynomials have been  possible up to now.

In section 5. we will introduce a new  procedure
for working with the
orthogonal and independent D-loop basis
which combines the above two alternatives
by constructing a suitable mapping of the
character functions on the D-loop basis
avoiding, however, 
the explicit handling of $SU(n)$ coupling coefficients.
In this framework, the Cayley-Hamilton relationship
in its general form is mainly used for
systematically computing certain norm relations.

Our procedure relies essentially on the 
following  observations:

1) The characteristic coupled cluster matrix elements
emerge as a byproduct when the 
(non-orthogonal and overcomplete) character
functions are set up systematically by an
iteration procedure.

2) The D-loop functions can - up to a 
normalization  factor - be 
uniquely characterized 
by the eigenvalue pattern of a  certain set
of commuting Casimir operators.

3) The matrix elements of these Casimir operators
are computable within the character functions
by the same methods which were used to set up
these functions.

Diagonalizing all necessary Casimir operators
in the space of character functions
yields then the mapping on the D-loop states.

This solves the problem of linear
dependencies among these functions by using
the eigenvalue patterns of the 
Casimir operators and by 
computing the relative norms of the dependent
eigenstates with the Cayley Hamilton relation.

The final calculations are in this way
reduced to a calculation of the
Kogut-Susskind eigenvalue problem
within the  D-loop basis expansion.

We hope that this will allow future
numerical Hamiltonian lattice QCD calculations
which may go to  higher order than
the previous attempts\cite{Llew, Guo,Luo, Mak}. 

Also, our procedure  yields a natural
truncation prescription for the
corresponding coupled cluster equations
because the D-loop basis is orthogonal
and unique.
 
Some details of a computational strategy are
described in section 6.

\section{Group Theoretical Structures}
We shall first give the definitions and notations for
the  $SU(n)$ lattice Yang-Mills theory,
especially its group theoretical content.

The general framework was given by 
Kogut and Susskind\cite{KoSu75}.
Accordingly, one has to define a Hilbert space ${\cal H}$
given by the set of ``top''
wave functions depending on $N$ link variables
\be
{\cal H} = \{\Psi(U_1,..U_N) \}
\ee
 where the quantities $U_l$
 ($l = 1,..,N$) are elements of the gauge group
$SU(n)$ and $N$ is the number of 
oriented links in a D-dimensional  lattice
(D is the number of space dimensions).

As in thermodynamics we shall work with
a finite volume, i. e.  with a finite 
lattice,  for definiteness. However,
our computational framework  allows one to take
  an infinite volume limit ($N \rightarrow \infty$)
at any later stage.

The scalar product is given by an N-fold
Haar measure integral.

The group theoretical nature of the link variables
$U_l$  gives as a natural 
orthogonal and complete basis of ${\cal H}$
all N-fold products
of $SU(n)$ D-functions, e.g. for $SU(2)$
we have the functions
\be
D^{j_1}_{m_1,m'_1}(U_1) D^{j_2}_{m_2,m'_2}(U_2)
 .. D^{j_N}_{m_N,m'_N}(U_N)
\ee

The group of (time independent)
local lattice gauge transformations is abstractly given by
\be 
G_{loc} = [SU(n)]^M
\ee
where $M$ is the number of {\em sites} of the lattice.

Elements of $G_{loc}$ are written as
$g = g(x)$ where $ x$ denotes any lattice site. 
A unitary representation
of $G_{loc}$ on ${\cal H}$ is then given by
\be
(\rho(g)\Psi)(U_1,..,U_N) = \Psi(U^g_1,..,U^g_N)
\ee
where the link variables are transformed like
parallel transporters:
\be
U_l^g  = g(x) U_l g^{-1}(x + \epsilon e_j)
\ee
if the link $l = (x, e_j)$
 connects the sites $x$ and $x + \epsilon e_j$
($\epsilon$ is the lattice spacing, $e_j$ is a
positive  unit vector
in $j$-direction).

The physical Hilbert space is defined by the subspace
of ${\cal H}$ corresponding to the trivial 
part of the decomposition of the
representation $\rho$, i.e. by the
gauge invariant states
\be
{\cal H}_{phys} = \{ \Psi \in {\cal H}|\; \rho(g) \Psi = \Psi
  \;\; \tf{for all g} \in G_{loc} \}
\ee
A systematic construction of a basis of
${\cal H}_{phys}$
generalizing Refs.\cite{Irv1,Guo} 
will be the main topic of this paper
and is 
described in section 4. 

We want to impose on this basis 
the classification of being
characterized by the irreducible
representations of the lattice
Euclidean group,
which is a strict symmetry group
of the lattice Kogut-Susskind Hamiltonian.
The lattice Euclidean group is a discrete remnant of the
standard continuum Euclidean group and
  is defined as follows: 

Let
\be
R^D_{latt} = \{ x = \epsilon \sum_{j=1}^D
 n_j e_j  |\;\; n_j = \tf{integer} \}
\ee
be the set of lattice sites of an infinite lattice.
The lattice translation group $G_{lt}$ is then isomorphic
to $R_{latt}^D$ and 
given by the mapping of $R^D_{latt}$
\be
x \rightarrow x+a
\ee
for any  $a \in R^D_{latt}$.
The lattice rotation group $G_{lr}$ is the restriction of
the group $O(D)$ leaving $R^D_{latt}$ invariant.
We call $G_{lr}$ the cubic group\cite{cubic}, it is discrete
and has 8 elements for $D=2$ and 48 elements for $D=3$.
The structure of this cubic group and its representations
are well known\cite{cubic}.
The Euclidean group $G_E$ is then the semidirect product
 $G_E = G_{lr} \otimes_s G_{lt}$  defined for $u = (R,a) \in G_E$ 
by the mapping of $R^D_{latt}$
\be
x \rightarrow ux = Rx + a 
\ee
Since the mappings $u$ may change the orientation
(i.e. it may be that $det(R) = -1$), the group
$G_E$ acts on the set of links with both
orientations. We use the notation
$\lambda = (l,\sigma), \sigma = \pm 1$
for these generalized links:

$(l,1)$ stands  for the links
with the originally chosen orientation,
i.e. they have the structure  
 $(l,1) = (x,e_j)$, connecting
$x$ to $ x + \epsilon e_j $
($x \in R^D_{latt}$) where  $e_j$ is $\em positive$.

 $(l,-1) = (x + \epsilon e_j,-e_j)$ connects then
$x + \epsilon e_j$ to $x$.

Writing $ \lambda = (x,c_j)$
for a general link, 
$c_j$ being a positive
or negative lattice unit  vector,
the action of $ u = (R,a) \in G_E$
is simply given by
\be
\lambda \rightarrow  u \lambda = ( ux,R c_j)
\ee

This allows us to define a unitary representation $T$ of the
lattice Euclidean group $G_E$ on the 
Kogut-Susskind wave functions
as a combination of 
the corresponding permutation of the link variables
and the mapping $ U \rightarrow U^{-1} $ if the
link in question is reoriented:
If $\Psi$ depends on the variables $U_{l_1},..,U_{l_r}$
and if we put $ u (l_\alpha,1) = (n_\alpha,\sigma_\alpha) $ 
($\alpha = 1,..,r$, $\sigma_\alpha = \pm 1$),
then $T(u)\Psi$ depends on the variables 
$U_{n_1},..,U_{n_r}$ and  we have
 \be
(T(u)  \Psi)(U_{n_1},..,U_{n_r}) = 
\Psi(U^{\sigma_1}_{l_1},..,U^{\sigma_r}_{l_r})\;.
\ee
As in the formal continuum limit, the operators
$T(u)$ commute 
with the Kogut-Susskind Hamiltonian
for all $u \in G_E$. 

We now  
construct projection operators
on subspaces of ${\cal H}$ or $ {\cal H}_{phys}$
corresponding to 
specific irreducible representations of $G_E$:

For translations we have a ``lattice momentum projection''
\be
\Pi_{lt}(p) = \sum_{a \in R^D_{latt}} \e{-i a_j p_j} T(R=1,a)
\ee
where $p \in R^D $ is restricted to the first Brillouin zone
($-\pi \leq \epsilon p_j \leq \pi$).

If $d^\Gamma_{\nu,\nu'}$ denote the D-functions for
the irreducible representations
(including inversions) of the cubic group
$G_{lr}$\cite{cubic}, a projection on "lattice
angular momentum''  $\Gamma$ is given by
\be
\Pi_{lr}(\Gamma;\nu,\nu') = 
\sum_{R \in G_{lr}} d^\Gamma_{\nu\nu'}(R) T(R,a=0)
\ee

By construction these definitions guarantee 
for any $\Psi \in {\cal H}$ the
characteristic relations
\bea
T(1,a) \Pi_{lt}(p)  \Psi &=& \e{i p_j a_j} \Pi_{lt}(p) \Psi
\\ \nonumber
T(R,0) \Pi_{lr}(\Gamma;\nu,\nu') \Psi & =& 
\sum_{\nu"} d^\Gamma_{\nu,\nu"}(R) \Pi_{lr} (\Gamma;\nu",\nu') \Psi
\eea

A combination of both projections yields states
with ``good'' momentum and angular momentum
in the sense that we have for
\be
\Psi^{\Gamma p}_{\nu\nu'} =
 \Pi_{lt}(p) \Pi_{lr}(\Gamma,\nu\nu') \Psi
\ee
the relations
\bea
T(1,a) \Psi^{\Gamma p}_{\nu\nu'} & = &
\e{i p_j a_j} \Psi^{\Gamma p}_{\nu\nu'}
\\ \nonumber
T(R,0) \Psi^{\Gamma, p}_{\nu\nu'} & = &
\sum_{\nu''} d^\Gamma_{\nu\nu''} \Psi^{\Gamma, Rp}_{\nu''\nu'}
\eea

The basic problem of a ``lattice Yang Mills theory'' is
then to find in ${\cal H}_{phys}$ (approximate) 
eigenfunctions  of the type $\Psi^{\Gamma p}_{\nu\nu'}$ of
 the Kogut-Susskind Hamiltonian\cite{KoSu75}
$H_{KS} = \frac{g^2}{2 \epsilon} H$ with
  \bea
    H &=& E_{l a} E_{l a} - x V
\\ \nonumber
     V & = & \sum_{\Box} \chi_{\Box}
\\ \nonumber
x & = & \frac{2}{g^4}
  \eea
where $g$ is the coupling constant
and $a$ is a colour index
($a=1,..,n^2-1$).
 Summation over
repeated indices is always assumed; $\Box$ labels the
plaquettes, and
$\chi_{\Box}$ is given by
  \be
    \chi_{\Box} := \tr 
U^{\sigma_1}_{l_{1}} U^{\sigma_2}_{l_{2}} 
U^{\sigma_3}_{l_{3}} U^{\sigma_4}_{l_{4}}
  \ee
when $\Box = (l_{1},\sigma_1, \ldots, l_{4},\sigma_4) $. The 
``colour-electric field operators''
$E_{l a}$ generate - in analogy to the standard momentum operator -
a left multiplication of group elements in the
arguments of  the wave functions. 
They are quantum operators conjugate to
the link operators $U_l$ 
obeying the commutation relations
  \be
    \sqbr{E_{l a} , U_{l'}}
                                 = \delta_{l l'} \lambda^{a} U_{l}
  \tf{ ,}
  \ee
where the SU(n) generators $\lambda^{a}$ are normalized according to
 $\tr \lambda^{a} \lambda^{b} = \delta^{ab} / 2$.

\newpage

\section{The exp S method
and intrinsic wave functions}

A motivation for the introduction of the coupled cluster or
exp~S method is given by the following considerations:

Given a Hamiltonian $H$, a standard and often 
successful method to get the approximate
spectrum of the  low lying energy states
is provided by the Lanczos approach:
Choose some trial state $\phi$ and diagonalize
H restricted to the finite-dimensional space
spanned by ($\phi, H\phi, H^2 \phi, .., H^n \phi$).
There are many cases where this gives reliable results
if $n$ is large enough.

However, for our lattice Yang Mills case, this procedure
is doomed to fail\cite{Hollenberg} because
we have here a situation analogous to  nuclear matter, for instance.
In the infinite volume limit ($N \rightarrow \infty$)
 -  where we want to
formulate our approach - the groundstate
energy $E_0$ and excitation
energies $ E - E_0$  of $H_{KS}$ have the behaviour
\bea
E_0 &\propto & N
\\ \nonumber
E - E_0 & \propto & 1
\eea
Also the groundstate wavefunction displays a
characteristic ``pathology'' 
in sense that its  norm (defined by the N-fold
Haar measure integral) has an essential singularity
for $N \rightarrow \infty$.  
Its precise structure will be given  below,
within perturbation theory it
is related to the appearance of disconnected
diagrams.

It has been known for a long time that this
difficulty is cured by 
rewriting the eigenvalue problem within the
exp~S framework (see Ref.\cite{Ripka} for the
standard many-body theory and Ref.\cite{Greensite} 
for the Kogut-Susskind theory).

For our case,
 the method consists of
introducing the  ans\"atze
\be
\Psi_0(U) = \exp{S(U)}
\ee
($U = (U_1,..,U_N) $)
for the ground state and
\be
 \Psi(U) = F(U) \exp{S(U)}
\ee
for excited states. 

The mentioned ``pathology'' of the ground state
consists then in the fact that we have the norm relation
 $|S|^2 \propto N$ for the function $S(U)$ appearing
in the exponent with respect to $\Psi_0$!

The validity of these volume dependencies is
related to a characteristic linked cluster structure
of $S(U)$ and $F(U)$, which follows
from rewriting the Schr\"odinger equation in terms
of these functions, resulting in the
non-linear equation
\be
 S_{\mu\mu} + S_\mu S_\mu 
   - x V =  E_0
\ee
for $S$ and in the linear equation
\bea
F_{\mu\mu} + 2 S_\mu F_\mu &=&  (E-E_0) F\;\;.
\eea
for the excitation operator $F$.

Here, we use the abbreviation
\be
\mu = (l,a)
\ee
and the notation
\be
f_\mu = \sqbr{E_{la},f}\;\;,\;\;
  f_{\mu\mu} = \sqbr{E_{la},\sqbr{E_{la},f}}
\ee
for any function $f(U)$.

Note that the ``coupled cluster 
equations'' (23) and (24) are still rigorous.

The linked cluster structure of
the functions $S(U)$ and $F(U)$
follows from the fact that they may be
expressed with the help of the projection
operators (12) and (13) as in (15) in terms of
``intrinsic'' functions.
These intrinsic functions are given by linked
clusters  and  are defined as follows.

Suppose that
$F\Psi_0$ describes a state with
Euclidean quantum numbers $(p,\Gamma,\nu,\nu')$
 (see eq. (15)). 
We then write $F$ and $S$ in the form
($S$ has to have trivial quantum numbers)
\bea
F(p,\Gamma,\nu,\nu') & = & \Pi_{lt}(p) 
\Pi_{lr}(\Gamma;\nu,\nu') F_{int}(p,\Gamma,\nu,\nu')   
\\ \nonumber
S &=& \Pi_0 S_{int}
\\ \nonumber
\Pi_0   & = & \Pi_{lt}(0) \Pi_{lr}(0,0,0)
\eea

If $F \Psi_0$ corresponds  
in the continuum limit to a {\em bound state}, we expect that
$F_{int}$ 
may be chosen to describe a {\em localized} state.
This is analogous to non-relativistic many-body theory
where bound states can be separated into square integrable
functions of the relative coordinates and an overall center
of mass motion, described here with projection operators.

In analogy  to nuclear matter for instance, the same 
localization holds true for the vacuum function
$S_{int}$ because correlations have a finite 
range.

The validity of these properties of the intrinsic
functions is seen below through the structure
of the expansion of these functions in terms of
a localized basis, i.e. a basis of {\em linked clusters}. 

We shall first characterize this basis  
through its general properties and then describe 
the concrete construction in section 4.

We call the basis
\be
\chi^\alpha(U_{l_1},..,U_{l_{m_\alpha}})
\;\;\alpha = 1,2,3,..
\ee
and impose the following conditions:

1) $\chi^\alpha$ should be gauge invariant.

2) $\chi^\alpha$ should be ``linked'', see 
section 4 for the precise definition.
A main consequence is that
$m_\alpha$ is finite for any $\alpha$, though
not limited.

3) $\chi^\alpha$ should be ``standardized'', i.e. the
equation
\be
T(u) \chi^\alpha = \lambda \chi^\beta \;\;(u\;\in\;G_E)
\ee
should only have solutions for $\alpha = \beta$.

4) $\chi^\alpha$ should be a strong coupling eigenfunction,
i.e. 
\be
\sum_{a} E_{l,a}E_{l,a} \chi^\alpha =
\epsilon_{\alpha,l} \chi^\alpha
\ee 

It will be convenient to specify $\chi^1$ as the ``plaquette
function''  by putting
\be
\Pi_0 \chi^1 = 4 (D-1) V
\ee
and to distinguish the constant function via
\be 
\chi^0 = 1\;\;.
\ee
$\chi^0$ fulfils the relation
\be
\Pi_0 \chi^0 = s_0 N \chi^0
\ee
where the symmetry factor $s_0 = |G_{lr}|$ is equal to 8,48 for 
$D = 2,3$, respectively.

Simplifying (27) by writing
$F = \Pi F_{int}$ and introducing the functions
${\tilde S}(S)$ and ${\tilde F}(S,F)$ by
\bea
(\Pi_0 S_{int})_\mu (\Pi_0 S_{int})_\mu &=& \Pi_0{\tilde S}
\\ \nonumber
(\Pi F_{int})_\mu (\Pi_0 S_{int})_\mu & = & \Pi {\tilde F}
\eea

the coupled cluster equations (23) and (24) can be rewritten
as
\bea
 (S_{int})_{\mu\mu} + {\tilde S} - \frac{x}{4(D-1)} 
\chi^1 & =&s_0 \frac{E_0}{s_0 N}
\\ \nonumber
 (F_{int})_{\mu\mu} + 2 {\tilde F} & =& (E-E_0)F_{int}
\eea

The ``linked cluster theorem'' for our lattice Yang-Mills theory
consists then in the statement that if (for $N \rightarrow \infty$)
we have the norm relation $|S_{int}|, |F_{int}| \propto 1$,
we have also  $|{\tilde S}|, |{\tilde F}| \propto 1$.

The norm relations for $S_{int}$ and $F_{int}$ 
are fulfilled because
of their localized nature, those for ${\tilde S}, {\tilde F}$
follow then as a result of the
fact that the ``derivative'' $S_\mu F_\mu$ selects
only linked clusters. 
We shall prove this structure by 
suitable expansions in terms of
the linked cluster basis (28).

Introducing the summation conventions
\bea
\sum_\alpha &=& \sum_{\alpha = 1,2,...}
\\ \nonumber
\sum_\alpha^{\prime} &=& \sum_{\alpha = 0,1,2,...}
\eea
the expansions of the intrinsic wavefunctions read
(specifying again the Euclidean quantum numbers)
\bea
S_{int}(U) &=& \sum_\alpha S_\alpha \chi^\alpha(U)
\\ \nonumber
F_{int}(p,\Gamma,\nu,\nu';U) &=& \sum_\alpha^{\prime} 
F_\alpha(p,\Gamma,\nu,\nu') \chi^\alpha(U)
\eea

The coupled cluster equations
 (23) and (24)  may then equivalently be formulated
as equations for the coefficients $S_\alpha$ 
and $F_\alpha(p,\Gamma,\nu,\nu')$ 

\bea
\epsilon_\alpha S_\alpha + \sum_{\beta,\gamma}  
C^{\beta\gamma}_\alpha S_\beta S_\gamma &=& 
\frac{x}{4(D-1)} \delta_{\alpha 1} 
           +\frac{E_0}{s_0 N} \delta_{\alpha 0}
\\ \nonumber 
    \epsilon_\alpha F_\alpha(p,\Gamma,\nu,\nu') \;+ & &
\\ \nonumber
  2 \sum_{\beta,\gamma,\nu_1}^{\prime} 
C^{\beta\gamma}_\alpha(p,\Gamma,\nu_1,\nu') 
S_\beta F_\gamma(p,\Gamma,\nu,\nu_1) &=& 
           (E-E_0) F_\alpha(p,\Gamma,\nu,\nu')
\\ \nonumber
\epsilon_\alpha &= & \sum_{l} \epsilon_{\alpha,l}
\eea

The crucial
quantities in these equations 
are the coupled cluster matrix elements
$C^{\beta\gamma}_\alpha$ 
and $C^{\beta\gamma}_\alpha(p,\Gamma,\nu_1,\nu')$
defining the expansion of the functions ${\tilde S}$
and ${\tilde F}$, respectively, 
which are obtained as follows:

Determine first the set of numbers $c^{\beta\gamma}_{\alpha u}$
($u \in G_E$) defined by
\be
\sum_{\mu,u\in G_E} 
\chi^\beta_\mu (T(u) \chi^\gamma)_\mu
= \sum_{\alpha,u}^{\prime}
 c^{\beta\gamma}_{\alpha,u} T(u)\chi^\alpha
\ee
This yields then
\bea
C^{\alpha \beta}_\gamma (p,\Gamma,\nu,\nu') &=&
\sum_{u=(R,a)} c^{\alpha\beta}_{\gamma,R,a} 
d^\Gamma_{\nu\nu'}(R^{-1}) \e{-i p_j a_j}
\\ \nonumber
C^{\alpha\beta}_\gamma &=& C^{\alpha \beta}_\gamma (0,0,0,0)
\eea
A proof hereof is given in the appendix. 
The linked cluster theorem guaranteeing the
correct volume dependencies of the relevant
quantities discussed above is now given by the
fact that,
due to the localized nature of the functions
$\chi^\alpha$, the r.h.s. sums in eqs. (39) and (40) 
run only over a finite number of terms. 

The coupled cluster equation (38)  also
 have the property
that approximate solutions generated by
truncations (see e.g. Refs.\cite{Llew,Guo})
display  correctly all relevant volume
dependencies.

The important task is now to compute
the matrix elements $c^{\beta\gamma}_{\alpha u}$,
for which one needs an efficient and
systematic way to set up and handle
the basis elements $\chi^\alpha$.

\section{Construction of the loop space basis}

In principle, an orthogonal basis of the type
$\chi^\alpha$ has been constructed in Ref.\cite{Irv1}:
for simplicity, we will formulate the method
for $SU(2)$ and $D=2$, but the generalization is
obvious, though technically more difficult.

Suppose $\chi^\alpha$ 
($\alpha$ fixed) depends on the link variables 
$(U_1,..,U_r)$. This basis function is then
- -  up to a normalization factor  -
{\em uniquely} characterized by the following set
of angular momenta: 

1) We have a set
 $ (J_1,..,J_r) $, i.e. one (half integer)
angular momentum for each link .

2) We have an
angular momentum 
$J_{ab} = J_{cd}$ for each quadruplet
of links $l_a,..,l_d$ forming a 4-point vertex
in the link pattern $(l_1,..,l_r)$. 
Here
the convention is that the links $ (l_a,l_b)$
are oriented such that they are going into,
and $(l_c,l_d)$ are leaving, the common site.

These angular momentum quantum numbers are
constrained by 
$J_a = J_b$ if $ (l_a,l_b) $ form a 2-point vertex
and by the condition that the coupling
$J_a + J_b \rightarrow J_c$ should be possible
if $(l_a,l_b,l_c)$ form a 3-point vertex.

For instance,  putting $r=7$ and choosing
the link pattern of Fig. 1, one
has just three angular momenta 
($J_1 (= J_2 = J_3), J_4 (= J_5 = J_6), J_7$) 
yielding the basis elements ($\alpha = (J_1,J_4,J_7)$)
\bea
\chi^\alpha &=& \sum_{M_1,..,M_{10}}
 D^{J_1}_{M_1,M_2}(U_1) D^{J_1}_{M_2,M_3}(U_2) 
 D^{J_1}_{M_3,M_8}(U_3^{-1}) 
\\ \nonumber
  & & \left( {{J_1, J_7, J_4} \atop {M_8,M_{10},M_4}} \right) 
  D^{J_{7}}_{M_7,M_{10}}(U_7)  
\left( {{J_1,J_7,J_4} \atop {M_1,M_7,M_9}} \right)
\\ \nonumber
& & D^{J_4}_{M_4,M_5}(U_4^{-1}) D^{J_4}_{M_5,M_6}(U_5^{-1} )   
  D^{J_4}_{M_6,M_9}(U_6)  
\eea
We call this orthogonal set of functions
the ``D-loop basis''.
 
In Ref.\cite{Irving} this framework was used for estimating
observables within the ELCE method, but higher order
calculations were limited by the 
necessity to handle an increasing number of $SU(n)$ couplings.
Also it should be mentioned that in Refs.\cite{Irv1,Irving}
the linked cluster form of the lattice Yang-Mills 
many-body problem was  taken into account
within a different computational framework.

An alternative for the construction of a basis
is related to an exp~S generalization of the
Lanczos idea and  was 
pursued in Refs.[6-10]
for trivial representations of the
Euclidean group. We will define this
method here in such a way that it allows 
the computation of arbitrary Euclidean
representations and also a transition
to the independent, orthogonal 
D-loop basis. This yields especially a
systematic way of eliminating
linear dependencies.

Starting with $\phi^{1,1} = \chi^1$ from (31) we
define ``character functions'' 
$\phi^{\delta,1},..,\phi^{\delta,n_\delta} $ 
by the iterative
condition that the following expansion
should  hold (we put $\phi^{0,1} = 1, n_0 = n_1 = 1$)
\bea
\phi^{\delta_1,k_1}_\mu (T(u)\phi^{\delta_2,k_2})_\mu &=&
\sum_{v \in G_E}
\sum_{0 \leq \delta \leq \delta_1 + \delta_2, k \leq n_\delta}
\zeta^{\delta_1 k_1 \delta_2 k_2}_{\delta k,u,v} T(v)\phi^{\delta,k}
\\ \nonumber
k_1 \;\leq \;n_{\delta_1}\;\; k_2 \; \leq \; n_{\delta_2} & &
\eea
Because of the ``derivative'' $\mu$, the functions
$\phi^{\delta,k}$ are by definition linked -
both terms on the l.h.s. have to have a common link
variable for a non-vanishing result - and they
also can be chosen to be standardized - i.e to obey
the condition (29) - because we included the 
Euclidean operator $T(v)$
on the r.h.s. of (42). 

The definition of the functions $\phi^{\delta,k}$
is made complete and unique  
by the condition that it should be just
a product of characters, i.e. for each $(\delta,k)$
there should exist a set of 
$L_1,..,L_r$
defined by the generalized links 
\be
L_j = (l_{j 1},\sigma_{j 1},..,l_{j {m_j}},\sigma_{j {m_j}})
\ee
such that
\be
\phi^{\delta,k} = \Pi_{j=1}^r 
tr(U^{\sigma_{j 1}}_{l_{j 1}}..U^{\sigma_{j {m_j}}}_{l_{j {m_j}}}) 
\ee
Hereby, for $SU(3)$, the loops $L_1,..,L_r$ 
should have all possible orientations
compatible with the standardization of $\phi^{\delta,k}$.
 For SU(2), however,
all loops should have the same (fixed) orientation
which is no loss of generality
because of the relation
$tr g = tr g^\dagger$ for $g \in SU(2)$.

The result of the l.h.s.  of (42) may be
expanded in such terms  because of an inductive
argument:  $\phi^{1,1}$ has the form (44).
Assuming the form (44) for the two terms of the
l.h.s. of (42) which we call,
specifying for simplicity only the
dependence on a certain (common) link
variable $U_l = V$,
$tr(AV)$ for the first term 
and $tr(BV^\sigma)$ for the second term 
 $ (\sigma = \pm 1)$, 
 the ``differentiation'' 
with respect to $ \mu = (l,a)$ 
may be evaluated using eq. (19) and 
 the standard property of the
$SU(n)$ generators $\lambda^a$:
\be
\sum_a \lambda^a_{ij}\lambda^a_{i'j'} = \frac{1}{2}
      (\delta_{ij'} \delta_{ji'} -
        \frac{1}{n}\delta_{ij}\delta_{i'j'})
\ee
yielding
\be
\sum_a (tr(AV))_{l,a}(tr(BV^\sigma))_{l,a} =
\sigma (\frac{1}{2} tr (V^\rho AV^\rho B) 
 -\frac{1}{2n} tr(AV ) tr(BV^\sigma))
\ee
where $\rho = (\sigma+1)/2$.
If the variable $V = U_l$  also occurs in
functions $A(U)$ or $ B(U)$, additional terms
arise on the r.h.s. due to the product rule  
of differentiation, but these terms will again display
a loop space structure of the same type.
The same happens if the first term has
the form $tr(AV^{-1})$.
In this sense, (46) describes the ``generic'' case.
 
Each character function $\phi^{\delta,k}$,
constructed in
this way, is uniquely characterized by the set 
of ``geometric'' loops indicated in (44). 
For $SU(2)$ and $D=2$, examples  up
to third order are given in Figs. 2 and 4.  
For this dimension and for SU(2), the number of
loop space functions is 
4, 16 for the orders $\delta = 2,3$, respectively.

\section{The D-Mapping} 
Of course, the system of loop space
functions $\phi^{\delta,k}$
is neither orthogonal nor linearly independent. For
SU(2), for instance, the number of independent functions
is known to be 
respectively 1,3,10 up to third order\cite{Guo}.
 
However, the construction  yields directly
the expansion
coefficients needed in (39) by taking in (42) the sum
over $u \in G_E$.

The main problem which remains is to select an independent subset in the space
of the functions $\phi^{\delta,k}$.

In Refs.\cite{Guo,Guo95} independent functions were
determined using besides
$tr g = tr g^\dagger$ the relation
$tr(gg') = tr g (tr g') - tr (g^\dagger g')$ for $g,g' \in SU(2)$.

The emerging functions were in general neither orthogonal
nor unique.

Within this paper, we propose a different
strategy, namely, 
 to  relate the functions  $\phi^{\delta,k}$ 
 - divided into convenient subsets - directly
to the orthogonal ``D-loop basis'' $\chi^\alpha$
by a
characteristic mapping, called {\em D-mapping}.

This allows us to do the final calculation,
i.e. to solve (approximately) eq.~(38),
with respect to the D-loop basis.
But, at the same time,
the crucial matrix elements $c^{\beta\gamma}_{\alpha,u}$
 may be computed in terms of the character functions
 using the D-mapping, while
avoiding any explicit
$SU(n)$-coupling or recoupling. We hope that
this simpler structure will finally allow
calculations of the type of Ref.\cite{Guo,Luo,Guo95}
to higher order and/or for $D=3$.

The construction of the (non-invertible) D-mapping
relies on the following structure 
of the D-loop basis already indicated in section 4.
For any $\chi^\alpha = 
\chi^\alpha(U_{l_1},..,U_{l_{r_\alpha}}) $
($\alpha$ fixed) there exists a maximal set of commuting
operators $(A_1,..,A_{M_{\alpha}})$ 
with the property that
\be
A_\lambda \chi^\alpha = a_\lambda \chi^\alpha
\ee
and such that the state $\chi^\alpha$ is
{\em uniquely} characterized by
the eigenvalues $(a_1,..,a_{M_{\alpha}})$.

In the case $SU(2)$ and   $D=2$,
these operators are apparently
given by a combination of the two sets
\be
\sum_a E_{\lambda a}E_{\lambda a} \;\;\;\lambda = 
        l_1,..,l_{r_\alpha}
\ee
and all operators 
\be
 \sum_a (E_{b a} + E_{c a})(E_{b a} + E_{c a})
\ee
which fulfil the condition that
$(b,c)$ are outgoing links of a four point vertex
in the link set related to $\chi^\alpha$.

The more general case may be extracted from Refs.\cite{Irv1,Irving}.
Important for our purpose is that
each $A_\lambda$ is a  Casimir operator
of the  local lattice gauge group $G_{loc}$,
i.e. it is a certain polynomial in the
operators $E_{l,a}$ of the type given above. 
For $SU(3)$,  two generalizations have 
to be taken into account:
The third order Casimir operators have to
be added to the set (47). In addition,
suitable permutation operators have to be
included if the $SU(3)$ Clebsch Gordan 
decompositions generalizing eq. (41)
have the property that the same irreducible
representations occurs several times.
(For $SU(2)$, an example for the definition
of such a permutation operator is given in the
appendix A3.)

The main point for the construction of the D-mapping
is that the evaluation of the operators 
$A_\lambda$ on the states $\phi^{\delta,k}$ 
can be done in precise  analogy to the
computation in eq. (46). The
important ingredient is again the relation (45)
and a corresponding third order  generalization
for SU(3) (see, e.g.\cite{Luscher}).
In other words, it is possible
to compute the matrix elements of $A_\lambda$
defined by
\be
A_\lambda \phi^{\delta,k} = 
     \sum_{\delta' \leq \delta, k'}(A_\lambda)^{\delta,k}_{\delta',k'}
                           \phi^{\delta',k'}
\ee
These matrices related to $A_\lambda$ are finite
$(\delta' \leq \delta)$ and stay small, in general.
This is because the terms on the r.h.s. have to
be consistent with the loop pattern of the
variables of the l.h.s.. 
Since link variables may be removed 
by $A_\lambda$ (see eq. (46) for $\sigma = -1$),
this consistency also allows  terms with 
removed variables on the r.h.s..
In the classification of the states 
as D-loop functions, this corresponds
to the possibility that one of the
link angular momenta may be zero. 
 The loop pattern of $\phi^{\delta,k}$
also determines  the choice of the possible  
operators $A_\lambda$ in (45).
More details are given in section 6 where
it is also shown that with the knowledge
of the matrices of $A_\lambda$ 
it is sufficient to work out (42)
without the derivatives.

Eq. (50) corresponds to an evaluation
of the  operators $A_\lambda$
 with respect to a non-orthogonal 
and overcomplete basis.
Because the $A_\lambda$ are hermitian
and commute with each other, they must
 nevertheless be simultaneously diagonalizable. 
Therefore, there must exist combinations
\be
\varphi^\gamma = \sum_{\delta,k} C^\gamma_{\delta,k} \phi^{\delta,k}
\ee
such that
\be
A_\lambda\varphi^\gamma = a_\lambda^\gamma \varphi^\gamma
\;\;\lambda = \gamma_1,..,\gamma_{r_\gamma}
\ee
The choice of the operators $A_\lambda$
is  determined by the loop pattern of the
variables occurring in $\varphi^\gamma$.

Any state $\varphi^\gamma$ is by construction
proportional to
a D-loop basis function characterized by
the eigenvalues 
$ a_{\gamma_1}^\gamma,..,a_{\gamma_r}^\gamma$.

Consequently, 
{\em the states $\varphi^\gamma$ are 
 - up to a normalization factor 
and a possible Euclidean mapping (11) - equal if and only if their
eigenvalue patterns are equal.}

Some subset of the
functions $\varphi^\gamma$ are  then
independent and orthogonal. They fulfill
all conditions of (28) and can be identified
with a certain subset of the $\chi^\alpha$.
The D-mapping is just the restriction of
(51) to such independent solutions and to
the computation of the relative normalization
factors for the dependent  states 
$\varphi^\gamma$.
Having determined the D-mapping, 
eqs. (51) and (42) contain all ingredients 
for the computation of the crucial matrix elements
(39).

\section{Computational strategy and examples}

We now describe the computational steps which -
put into the language of a suitable computer program -
would lead to the possibility to determine
approximate glueball spectra. We shall 
elucidate these steps by some examples
of low order for $D=2$.

\subsubsection*{1) Set up the character functions.}

In order to minimize the computational effort,
we propose to divide the character functions
into subsets of the following type.

Introduce first the set of ``generic'' functions
\be
\Lambda_G^{\delta,k}\;\;\delta = 1,2,..\;\; ; \;k=1,..,n_\delta
\ee
of $\delta$-fold {\em linked,
standardized plaquette products}.

For $SU(2)$, all plaquettes should have the
{\em same orientation}.

For $D=2$, we have for $SU(2)$ $n_\delta = 1,2,4$, 
for $SU(3)$
$n_\delta = 1,4,12$ up to $\delta = 3$.
 Fig. 2 gives the
corresponding loop patterns for $SU(2)$.

The relevance of this set of functions is 
two-fold:

i) They determine the possible elements of the
D-loop basis occurring up order $\delta$.
They are given by the link patterns
of the generic set and the
coupling rules of as many fundamental
representations (and its adjoint) as 
there exist  common links.
Hereby, a double counting with lower order
states has to be avoided.
Fig. 3  exemplifies the related elements of the
D-loop  basis for $SU(2)$ and D=2 up to 
$\delta = 3$.

ii)  The set (53) is 
``generic'' because
each element  defines
a characteristic subspace 
given by functions
\be
\Lambda^{\delta,k,\nu} \;\; ;\; \nu = 1,..,M(\delta,k)
\ee
which is left
invariant under the action of
any  Casimir operator of the lattice gauge
group. Hereby, of course, only a finite number
Casimir operators is relevant for any given
$(\delta,k)$. The set of character functions
$\phi^{\delta,\kappa}$ 
is contained in the set (54),
(i.e. for a given $\delta$ there exists 
for each $\kappa$ a pair $(k(\kappa),\nu(\kappa))$ 
such that that 
$\phi^{\delta,\kappa} = \Lambda^{\delta,k(\kappa),\nu(\kappa)})$)
an explicit construction is not necessary
because it is more convenient to work
with (54).

\subsubsection*{2) Compute the Casimir operator matrices.}
 
The set (54) is generated by applying the
relevant Casimir operators on (53)
yielding the matrix elements (50) as a 
system of - in  general small - 
submatrices
\be
A_\lambda \Lambda^{\delta,k,\nu} = \sum_{\nu' = 1}^{M(\delta,k)}
       A^{\nu}_{\nu'}(\lambda,\delta,k) \Lambda^{\delta,k,\nu'} 
\ee
Applying the product rule and (46) for the evaluation
of the Casimir operators yields for the 
loop structure of the subspaces
(54) ($\delta,k$ fixed)
the simple geometrical condition that
they are generated from the plaquette systems (54)
by ``cutting and glueing'' doubly occuring links.
Up to third order,  the related $SU(2)$ loop structures
for $D=2$ are given in Fig. 4, some
examples of the corresponding
 Casimir operator matrices   
are presented in the Appendix A2.

Note that the sets (54) may also contain elements
of lower order if they  occur during the cutting
and glueing procedure.
Also a standardization is
not done. This is convenient
since this makes the ``Casimir matrices''
(55) especially simple.

\subsubsection*{3) Fix the D-Mapping.}

The next step is the diagonalization of the
Casimir matrices (56) giving eigenfunctions
\be
\varphi^{\nu}(\delta,k) = \sum_{\nu'} 
 C^\nu_{\nu'}(\delta,k) \Lambda^{\delta,k,\nu'}
\ee
obeying
\be
A_\lambda \varphi^{\nu}(\delta,k) = a_\lambda (\nu,\delta,k)
                     \varphi^{\nu}(\delta,k)
\ee

For the construction of the D-mapping one first may put
with a suitable enumeration $\bar{\alpha}(\nu,\delta,k)$
\be 
\varphi^{\nu}(\delta,k) = N(\nu,\delta,k) T(\bar{u}(\nu,\delta,k))
       \chi^{\bar{\alpha}(\nu,\delta,k)}\;.
\ee
Hereby,  linear dependencies are eliminated by  the
identification prescription
\be
\bar{\alpha}(\nu,\delta,k) = \bar{\alpha}(\nu',\delta',k')
\Leftrightarrow
a_\lambda (\nu,\delta,k) = a_\lambda (\nu',\delta',k')
\;\; for\; all \; \lambda \;\;.
\ee

Of course, equality of the eigenvalue patterns
guarantees the equality of the corresponding eigenfunctions
only up to a (non-zero) factor $N(\nu,\delta,k)$
and up to a Euclidean transformation $ T(\bar{u}(\nu,\delta,k))$

For the computation of
the normalization factors $N(\nu,\delta,k)$ we observe
that within our exp~S framework
(including a possible truncation) it is not
necessary to work with basis states which
are normalized to one. Hence only the relative
factors 
are needed, i.e. we may put $N(\nu,\delta,k) = 1$
if the D-loop function $\chi^{\bar{\alpha}(\nu,\delta,k)}$
occurs for the first time when increasing the order $\delta$.
Also we may set for this first case $ \bar{u}(\nu,\delta,k) = 1$.
As a result we may find for each quantum number $\alpha$
an eigenfunction (56) characterized by 
$(\bar{\delta}(\alpha),\bar{\nu}(\alpha),\bar{k}(\alpha))$
defining an expansion of the elements of
the D-loop basis in terms of the character functions
\be
\chi^\alpha = \sum_{\nu'} 
 C^{\bar{\nu}(\alpha)}_{\nu'}
(\bar{\delta}(\alpha),\bar{k}(\alpha))
\Lambda^{\bar{\delta}(\alpha),\bar{k}(\alpha),\nu'}\;\;.
\ee   

For each $ (\delta,k)$, the matrices
$ C^\nu_{\nu'}(\delta,k)$ may be inverted,
yielding with (58) 
the inverse mapping
\be
\Lambda^{\delta,k,\nu} = \sum_{\nu'}
    D^\nu_{\nu'}(\delta,k) N(\nu',\delta,k)
T(\bar{u}(\nu',\delta,k))\chi^{\bar{\alpha}(\nu',\delta,k)}\;\;.  
\ee
Equation (61) - together with the
inversion (60) - constitutes the D-mapping in a form 
which is sufficient for the computation of
the coupled cluster matrix elements (39).

We still have to give a recipe to compute
the normalization
factors $N(\nu,\delta,k)$. In principle, they could
be determined by evaluating 
Haar measure integrals. This can be avoided, however,
by rewriting the states $\varphi^{\nu}(\delta,k)$
in a (up to the normalization factors) unique
form by using the usual procedure of eliminating
linear dependences via the Cayley Hamilton relations
\bea
tr g^\dagger &=& tr g\;\; 
\\ 
g^2 &=& g tr g - 1 \;\;\;(g \in SU(2))
\eea
and
\bea
  tr g^2 &=& (tr g)^2 - 2 tr g^\dagger 
\\ 
g^3 &=& g^2 tr g - 
 g tr g^\dagger + 1 \;\;(g \in SU(3))
\eea
This allows us to introduce a standardization 
of the functions $\Lambda^{\delta,k,\nu}$
by eliminating
for $SU(2)$ ($SU(3)$) 
all structures of the type $tr g^n g'$ with
$n \geq 2$ ($n \geq 3$).
For this purpose, eq. (61) respectively (63) have to
be iterated yielding formulas of the type
$g^n = a g^2 + b g + c$ where 
 $a, b,c$ are polynomials
in $tr g $ and $tr g^\dagger$. 
(For $SU(2)$, $a = 0$ and $b,c$ become
polynomials in $tr g$ only.)

For $SU(3)$ 
and $n = 2 $, also terms of the type
$tr g^2$ in (44) may be standardized 
with the help of (64).

For the examples 
where the Casimir matrices (55) 
are computed, we give in the
Appendix A2 also a
 construction of the corresponding part of the D-mapping.

\subsubsection*{4) The incorporation of the Euclidean group.}

For the computation of the matrix elements (39)
it is sufficient to work out (42) disregarding
the derivatives, i.e. to determine
the coefficients $\eta^{\gamma_1,\gamma_2}_{\gamma_3;u,v}$
($u,v \in G_E$) given by
\be
\Lambda^{\gamma_1} T(u) \Lambda^{\gamma_2} = \sum_{\gamma_3,v}
\eta^{\gamma_1,\gamma_2}_{\gamma_3;u,v} T(v) \Lambda^{\gamma_3}
\ee
where we introduced the abbreviation $ (\delta,k,\nu)= \gamma$.
Here, the character functions  
$\Lambda^{\gamma_1}$ and $T(u) \Lambda^{\gamma_2}$
have to fulfil the restriction that they are {\em linked},
i.e. they should have a common link variable.
Also the trivial function $\Lambda^{0,1,1} = \chi^0 $
should be left out. 

The following structures simplify the determination
of these $\eta$-coefficients:

a) There is only one 
non-vanishing term on the r.h.s. of (66). If 
 $\eta^{\gamma_1,\gamma_2}_{\gamma_3;u,v}$
is non-vanishing, it
 is equal to one. In this case we call the 
corresponding states
$\Lambda^{\gamma_1},\Lambda^{\gamma_2},\Lambda^{\gamma_3}$
{\em non-trivially connected}.

b) For each triple  
$\Lambda^{\gamma_1},\Lambda^{\gamma_2},\Lambda^{\gamma_3}$
of  non-trivially connected character functions
we have a characteristic set of Euclidean
group elements 
$u_\lambda,v_\lambda$ 
such that
\be
\Lambda^{\gamma_1} T(u_\lambda) \Lambda^{\gamma_2}
 = T(v_\lambda) \Lambda^{\gamma_3}
\; ; \; \lambda = 1,..,n(\gamma_1,\gamma_2,\gamma_3)
\ee

The determination of these elements $u_\lambda,v_\lambda$ 
is now simplified by the following structure:
Suppose we have found all solutions
$u_\lambda,v_\lambda$ 
for a non-trivially connected triple of {\em generic functions}
\be
\Lambda^{\delta_1,k_1}_G T(u_\lambda)
\Lambda^{\delta_2,k_2}_G = T(v_\lambda) \Lambda^{\delta_1,k_1}_G 
\ee
If $v_\lambda$ is suitably chosen, we have then
for each $\Lambda^{\delta_1,k_1,\kappa_1}$ and
$\Lambda^{\delta_2,k_2,\kappa_2}$ a function
$\Lambda^{\delta_3,k_3,\kappa_3}$ so that they are
non-trivially connected with the same set 
of Euclidean group elements as in (67)
and this exhausts all possibilities. 

Given the generic functions
of the r.h.s. of (68) and 
$\kappa_1,\kappa_2$,
the third character function $\phi^{\delta_3,k_3,\kappa_3}$ 
is then
determined by finding {\em just one} pair $(uv)$
solving (65).

Up to third order $\delta_3 = 3$,
a full computation of all $\eta$-coefficients in 
presented in Appendix A4.

\subsubsection*{6) The computation of 
the c-coefficients in  (39).}

Having solved the ``combinatorial'' problem of
determining the coefficients of (66), 
one may compute the quantities 
$c^{\alpha_1,\alpha_2}_{\alpha,u} $
by writing (66) in terms of the orthogonal
and independent basis $\chi^\alpha$ with
the help of (60,61) and by applying the
(Euclidean invariant)
``total Casimir operator'' $\sum_{la} E_{la}E_{la}$
on both sides of the emerging equation.
Writing $\bar{\gamma}(\alpha) = (\bar{\delta}(\alpha),
\bar{k}(\alpha),\bar{\nu}(\alpha))$, 
we obtain as final result for the crucial
coupled cluster matrix elements (39)
\bea
c^{\alpha_1,\alpha_2}_{\alpha_3,u} &=&
[\sum_{l} (\epsilon_{\alpha_1,l} + \epsilon_{\alpha_2,l}
 - \epsilon_{\alpha_3,l})]
\sum_{v \in G_E} \sum_{\gamma_1,\gamma_2,\gamma_3}
\\ \nonumber
& &\sum_{\gamma_4\; with\; \bar{\alpha}(\gamma_4) = \alpha_3 }
 C^{\bar{\gamma}(\alpha_1)}_{\gamma_1}
C^{\bar{\gamma}(\alpha_2)}_{\gamma_2}
N(\gamma_4)\eta^{\gamma_1,\gamma_2}_
{\gamma_3;
u, v (\bar{u}(\gamma_4))^{-1}}
D^{\gamma_3}_{\gamma_4}
\eea

\section{Discussion and conclusion}

The coupled cluster formulation of Hamiltonian
lattice QCD needs an efficient method to deal
with suitable basis systems of loop space
functions. Within this paper we have demonstrated
that it possible to combine the merits of
a D-function basis, used within the ELCE
framework\cite{Irving} with those of 
the character sets used within recent
coupled cluster attempts\cite{Guo,Llew}
without facing the respective deficiencies.

The merits are the orthogonality
of the basis in the first case, the
close relation to the Lanczos method
and the easy computability of the
coupled cluster matrix elements in the second case.
 
The deficiencies are the need of handling
too many $SU(n)$ recoupling coefficients
for the computation of the Hamiltonian
matrix elements  when using D-functions,
the non-orthogonality and linear dependence
of the states when using the character functions.

Our combination is based upon the simple
idea that the D-function basis may be
characterized by the quantum numbers of
a complete set of commuting operators.
These operators are the Casimir operators
of the local lattice gauge group 
(for the gauge group $SU(3)$, also certain
permutation operator have to be included)
and our method 
relies on the fact that these commuting operators
(where only a finite set is relevant for any
specific case) may be evaluated as finite
matrices with respect to the character functions.
This allows the construction of a systematic
mapping between the two frameworks.

Invoking the lattice Euclidean symmetry
of the regularized gauge field theory
and  
systematizing the action of this symmetry group,
we were also able to formulate
the coupled cluster
lattice Hamiltonian eigenvalue problem
for eigenstates with arbitrary lattice
momentum and lattice angular momentum.
The whole formulation may be done
in the infinite volume limit.  

For any concrete calculation of the spectrum,
a truncation prescription has to be defined.
This point has been 
much in dispute\cite{Llew,Guo,Luo,Guo95}
because previously one had to make a (non-unique) choice
of independent functions from 
the non-orthogonal set of character functions.
Within our method we have
a more natural definition because
the orthogonal D-function basis is uniquely determined.

We want to stress that any truncated
coupled cluster calculation will have
the same limitations as any (finite volume)
standard lattice Monte Carlo computation,
namely that at best one has to hope
for a scaling window indicating consistency
with respect to the predicted renormalization group
structure which has to be displayed by any
observable when approaching the continuum limit.
(This structure is still unclear within Ref.\cite{Guo95} which gives
the ``best'' coupled cluster results up to now.) 

The reason for this expected scaling window in given by 
the fact that the truncation which has to be
defined with respect to an expansion of
the {\em intrinsic} wave functions of the
vacuum and of the hadron, necessarily limits
the possible lattice volume over which the physical
states may extend. Consequently,
when the physical lattice scale is set
by choosing the coupling $g$, the
method has to break down 
when the physical lattice volume,
given by the truncation - or by
the number of lattice points
in the
standard lattice Monte Carlo case - 
becomes smaller than the size of 
the hadron.

First attempts at doing concrete numerical
calculations within the reported framework
are on the way and will be reported in the
future\cite{Weichmann}.

Finally we want to mention that our
computational framework may, in principle,
be easily extended to  include
Fermions. Especially, a formulation
for Wilson Fermions within a quenched
approximation yields
equations whose treatment appear to be
no more complicated than that for
glueballs. Details of this structure
will be reported elsewhere. 
\subsubsection*{Acknowledgements}

D. S. wants to appreciate
illuminating discussions
with H. Kr\"oger, H. Petry, N. Scheu, P. Schuck
and C. Weichmann.  
Financial support of the Deutsche Forschungsgemeinschaft
is also gratefully acknowledged.

\section*{ Appendix}
\subsubsection*{A1. Proof of equation (37)}
Introducing the abbreviation
\be
D^{p,\Gamma}_{\nu,\nu'}(R,a) = d^\Gamma_{\nu\nu'}(R) \e{i p_j a_j}
\ee
the expansion of $ F_\mu S_\mu $ according to (27) and (34)
yields the relevant terms
\bea
& &(\Pi_{lt}(p) \Pi_{lr}(\Gamma;\nu,\nu') \chi^\alpha)_\mu 
(\Pi_0 \chi^\beta)_\mu 
\\ \nonumber
&=&\sum_{u_1,u_2 \in G_E} 
D^{p,\Gamma}_{\nu,\nu'}(u_1) (T(u_1) \chi^\alpha)_\mu
(T(u_2) \chi^\beta)_\mu
\\ \nonumber
&=& \sum_{u_1,u_2 \in G_E} D^{p,\Gamma}_{\nu,\nu'}(u_1) 
(T(u_1) \chi^\alpha)_\mu
(T(u_1) T(u_1^{-1} u_2) \chi^\beta)_\mu
\\ \nonumber
&=& \sum_{u_1} D^{p,\Gamma}_{\nu,\nu'}(u_1) T(u_1)
[\sum_{u} \chi^\alpha_\mu (T(u)\chi^\beta)_\mu] 
\\ \nonumber
&=& \sum_{u_1} D^{p,\Gamma}_{\nu,\nu'}(u_1) T(u_1)
\sum_{u,\gamma} c^{\alpha,\beta}_{\gamma,u} T(u)\chi^\gamma
\\ \nonumber
&=& \sum_{u,\gamma} c^{\alpha,\beta}_{\gamma,u} 
 \sum_{u_1} D^{p,\Gamma}_{\nu,\nu_1}(u_1 u) 
D^{p,\Gamma}_{\nu_1,\nu'}(u^{-1}) T(u_1 u) \chi^\gamma
\\ \nonumber
&=& 
\sum_{\nu_1}\Pi_{lt}(p) \Pi_{lr}(\Gamma;\nu,\nu_1)
\sum_{u,\gamma} c^{\alpha,\beta}_{\gamma,u} 
D^{p,\Gamma}_{\nu_1,\nu'}(u^{-1})  \chi^\gamma
\eea
Here we assumed (39) and we arrived at the result (40).

\subsubsection*{A2. Examples for the D-mapping}
Within this appendix we will give
the construction of ``Casimir matrices''
(55) for $SU(2)$ and $D = 2$ 
for some typical cases
of subspaces $(\delta,k)$ taken from in Fig. 4.
Subsequently, we will present the corresponding
part of the D-mapping resulting from 
a diagonalization.

Note that also the D-loop basis is characterized
by the same quantum numbers $(\delta,k)$, see Fig. 3.

\subsubsection*{($\delta$,k) = (2,1)}
Here, two parallel links with the same orientation
occur. If a given link of this type is denoted by $U$,
this case yields for the corresponding
Casimir operator (48) a $ 2 \times 2 $ matrix
with repect to  states of the type
\bea
\Lambda^1 &=& tr (BU) tr (CU)
\\ \nonumber
\Lambda^2 &=& tr (BU CU)
\eea
For $(\delta,k) = (2,1)$, we have $B=C$ and U may
be any of the four links. A more general case
is e.g. given by $U=U_2$ for $(\delta,k) = (3,2)$.
(For enumeration see Fig 3).
The evaluation of $A = E_a E_a$ yields 
\bea
A \Lambda^1 &=& \Lambda^1 + \Lambda^2
\\ \nonumber
A \Lambda^2 &=& \Lambda^1 + \Lambda^2
\eea
The diagonalization gives the Casimir spectrum
as an example of eqs. (56) and (57)
\bea
\varphi^1 &=&\Lambda^1 + \Lambda^2  \;\; ; \;\; a(1) = 2
\\ \nonumber
\varphi^2 &=&\Lambda^1 - \Lambda^2  \;\; ; \;\; a(1) = 0
\eea
For $A=B$ one may eliminate  
in the state $\varphi^2$ the function $(AU)^2$ via (61)
yielding the identifications 
with the D-loop functions (see Fig. 3 for notations)
\bea
\varphi^1 &=& \chi^{2,1}
\\ \nonumber
\varphi^2 &=& 2 = 2\chi^0
\eea
Note that we put the normalization factor $N(\nu,\delta,k)$
equal to one when the corresponding basis state
$\chi^\alpha$ occurs for the first time.
Within our examples, the Euclidean mapping
$T(u)$ in (58) is the identity in most cases.
 
With the inversion
\bea
\Lambda^1 &=& \frac{1}{2} \chi^{2,1} + \chi^0
\\ \nonumber
\Lambda^2 &=& \frac{1}{2} \chi^{2,1} - \chi^0
\eea
these formulas give the  D-mapping relevant for
the subspace $(\delta,k) = (2,1)$.

\subsubsection*{($\delta$,k) = (2,2)}
This  case is of the type
\bea
\Lambda^1 &=& tr (BU) tr (U^\dagger C)
\\ \nonumber
\Lambda^2 &=& tr (B C)
\eea
yielding
\bea
A \Lambda^1 &=& 2 \Lambda^1 - \Lambda^2
\\ \nonumber
A \Lambda^2 &=& 0 
\eea
A diagonalization of (80) gives
\bea
\varphi^1 &=&2\Lambda^1 - \Lambda^2  \;\;; \;\; a(1) = 2
\\ \nonumber
\varphi^2 &=&\Lambda^2  \;\;\;\;\;\; ;\;\; a(1) = 0
\eea
With the identification
\bea
\varphi^1 &=& \chi^{2,2,1}
\\ \nonumber
\varphi^2 &=&  \chi^{2,2,2}
\eea
and the inversion
\bea
\Lambda^1 &=& \frac{1}{2} (\chi^{2,2,1} + \chi^{2,2,2})
\\ \nonumber
\Lambda^2 &=& \chi^{2,2,2}
\eea
this defines all  the D-mapping ingredients for
$(\delta,k) = (2,2)$.

\subsubsection*{($\delta$,k) = (3,1)}

Here one has to deal with a three-dimensional
subpace given by
\bea
\Lambda^1 &=& (tr g)^3
\\ \nonumber
\Lambda^2 &=& tr g\; tr g^2
\\ \nonumber
\Lambda^3 &=& tr g^3
\eea
with $ g = BU $. The product rule yields
for the evaluation of the Casimir operator
\bea
A \Lambda^1 &=& \frac{3}{4} \Lambda^1 +3 \Lambda^2
\\ \nonumber
A \Lambda^2 &=& \Lambda^1 +\frac{3}{4} \Lambda^2 + 2\Lambda^3
\\ \nonumber
A \Lambda^3 &=& 3\Lambda^2 + \frac{3}{4} \Lambda^3
\eea
The eigenvectors and eigenvalues are
\bea 
\varphi^1 &=& \Lambda^1 + 3 \Lambda^2 + 2 \Lambda^3  
\;\;\;; \;\; a(1) = \frac{15}{4}
\\ \nonumber
\varphi^2 &=& \Lambda^1 - \Lambda^3  
\;\;\;\;\;\; ; \;\; a(2) = \frac{3}{4}
\\ \nonumber
\varphi^3 &=& \Lambda^1 - 3 \Lambda^2 + 2 \Lambda^3  
\;\;\; ; \;\;a(3) = -\frac{9}{4}
\\ \nonumber
\eea
Using $tr g^3 = (tr g)^3 - 3 tr g $ obtained from (63),
this yields the identifications
\bea
\varphi^1 &=& \chi^{3,1}
\\ \nonumber
\varphi^2 &=& 3\chi^{1,1}
\eea
Since the eigenvalue $a(3)$ is negative, we must have  
\be
\varphi^3 = 0
\ee
yielding the linear dependence relation
\be
 \Lambda^2 = \frac{1}{3}(\Lambda^1 + 2 \Lambda^3) = (tr g)^3 -  2 tr g
\ee
which is just the result for $\Lambda^2$ when eliminating
$g^2$ by (61).
The inversion (64) now reads
\bea
\Lambda^1 &=& \frac{1}{6} \chi^{3,1} + 2 \chi^{1,1}
\\ \nonumber
\Lambda^2 &=& \frac{1}{6} \chi^{3,1} 
\\ \nonumber
\Lambda^3 &=& \frac{1}{6} \chi^{3,1} - \chi^{1,1}
\\ \nonumber
\eea

\subsubsection*{($\delta$,k) = (3,4)}
This case is interesting because it involves a 4-point
vertex, the dimension of the subspace (54) is 5.
Calling the doubly occuring link variables $U_1,U_2$
the generating character states are of the type
\bea
\Lambda^1 &=& tr (BU_1)\; tr (U_1^\dagger C U_2)\; tr (U_2^\dagger D)
\\ \nonumber
\Lambda^2 &=& tr (B C U_2)\; tr (U_2^\dagger D)
\\ \nonumber
\Lambda^3 &=& tr (BU_1)\; tr (U_1^\dagger C D)
\\ \nonumber
\Lambda^4 &=& tr (B C D)
\\ \nonumber
\Lambda^5 &=& tr (U_1^\dagger C U_2)\; tr (B U_1 U_2^\dagger D)
\eea
Now we have three relevant Casimir operators, $A_1$
and $A_2$ as before of type (48) and
$A_3 = (E_{1a}+E_{2a})(E_{1a}+E_{2a}) - A_1 - A_2
= 2 E_{1a} E_{2a} $
which is of the type (49).
The related Casimir matrices are given by
\bea
A_1 \Lambda^1 &=& 2 \Lambda^1 - \Lambda^2
\\ \nonumber
A_1 \Lambda^2 &=& 0
\\ \nonumber
A_1 \Lambda^3 &=& 2\Lambda^3 -  \Lambda^4
\\ \nonumber
A_1 \Lambda^4 &=& 0
\\ \nonumber
A_1 \Lambda^5 &=& -\Lambda^4 +  2 \Lambda^5
\eea
\bea
A_2 \Lambda^1 &=& 2 \Lambda^1 - \Lambda^3
\\ \nonumber
A_2 \Lambda^2 &=& 2\Lambda^2 -  \Lambda^4
\\ \nonumber
A_2 \Lambda^3 &=& 0
\\ \nonumber
A_2 \Lambda^4 &=& 0
\\ \nonumber
A_2 \Lambda^5 &=& -\Lambda^4 +  2 \Lambda^5
\eea
\bea
A_3 \Lambda^1 &=& - 2 \Lambda^1 +\Lambda^2 + \Lambda^3 -\Lambda^5
\\ \nonumber
A_3 \Lambda^2 &=& 0
\\ \nonumber
A_3 \Lambda^3 &=& 0
\\ \nonumber
A_3 \Lambda^4 &=& 0
\\ \nonumber
A_3 \Lambda^5 &=& 2\Lambda^4 - 4 \Lambda^5
\eea
The simultaneous eigenfunctions are 
\bea
\varphi^3 &=&-2\Lambda^1 + \Lambda^2 +\Lambda^3
 -\Lambda^4+\Lambda^5  \;\; ;\;\; a_1(1) = a_2(1) = 2 \;\;a_3(1) = -2
\nonumber \\ \nonumber
\varphi^4 &=&-2\Lambda^5 +\Lambda^4 \;\;\;\;\;\;\; ;\;\;
a_1(2) = a_2(2) = 2 \;\;a_3(2) = -4
\\ \nonumber
\varphi^2 &=&-2\Lambda^2 +\Lambda^4 \;\;\;\;\;\;\;\; ;\;\; 
a_1(3)= 0\;\;  a_2(3) = 2 \;\;a_3(3) =0
\\ \nonumber
\varphi^5 &=&-2\Lambda^3 +\Lambda^4 \;\;\;\;\;\;\;\; ;\;\; 
a_1(4) = 2\;\; a_2(4) = 0 \;\;a_3(4) = 0
\\ 
\varphi^1 &=&\Lambda^4 \;\;\;\;\;\;\;\;\;\;\; ;\;\; 
a_1(5) = a_2(5) = a_3(5) = 0
\eea
with the identifications 
\bea
\varphi^\nu &=& \chi^{3,4,\nu} \;\;\nu = 1,2,3,4
\\ \nonumber
\varphi^5 & =& T(P)\chi^{3,4,2}
\eea
where $P$ describes the reflection
(parity transformation) defined in equation (96).

The inversion (64) reads
\bea
\Lambda^1 &=& \frac{1}{4}(-2\chi^{3,4,3} +\chi^{3,4,1}
 - \chi^{3,4,2} - T(P) \chi^{3,4,2} - \chi^{3,4,4})
\\ \nonumber
\Lambda^2 &=& \frac{1}{2}(\chi^{3,4,1} -\chi^{3,4,2})
\\ \nonumber
\Lambda^3 &=& -\frac{1}{2}(T(P)\chi^{3,4,2} -\chi^{3,4,1})
\\ \nonumber
\Lambda^4 &=& \chi^{3,4,1}
\\ \nonumber
\Lambda^5 &=& \frac{1}{2}(\chi^{3,4,1} - \chi^{3,4,4})
\eea

\subsubsection*{A3. The ``local action'' of the permutation group}
We explain this structure for the 
``typical'' example $(\delta,k) = (3,4)$
discussed in the Appendix A2, 
a generalization for general cases
is straightforward, but will not be displayed within
this paper.

For $SU(2)$, first the equivalence of the fundamental
representation and its adjoint has to be invoced
by introducing the skewsymmetric $2 \times 2$ matrix
\begin{displaymath}
{\epsilon} = \left( \begin{array}{ccc}
0 & 1 \\
-1 & 0 \end{array} \right)
\end{displaymath}
which has the property
that $ \epsilon g^{-1} \epsilon^{-1} = \tilde{g} $
for any $g \in SU(2)$.
Consequenty, the modified link variable $\bar{U} = U\epsilon$
obeys instead of (5)
the ``tensor product'' transformation rule
(written in terms of matrix elements)
\be
(\bar{U}_l^g)_{jk} = g(x)_{jj'} g(x + e_j)_{kk'} (\bar{U}_l)_{j'k'} 
\ee
Introducing  the ``modified loop group elements''
specified according to  the common
(four point) lattice site of our example (see eq.(89))
\bea
\alpha & = & BU_1\epsilon
\\ \nonumber
\beta & = & U_1^\dagger D U_2 \epsilon
\\ \nonumber
\gamma & = & U_2^\dagger D\epsilon
\\ \nonumber
\eea
the character states (89) may be written as
\be
\Lambda^\nu = d(\nu)^{j_1 j_2 j_3 j_4 j_4 j_6}
  \alpha_{j_1 j_2} \beta_{j_3 j_4} \gamma_{j_5 j_6}
\ee
Each d-coefficient couples the 
tensor product of the six fundamental representations
- - defined ``locally'' corresponding to the chosen 
 the four point vertex  - to the trivial
representation, i.e., introducing the
related six angular momentum operators
${\bf s}(r), r = 1,..,6$ 
($s_j$ are the Pauli spin matrices in our case)
the states (98)
obey
\bea
{\bf s}_{total} \Lambda^\nu &=& 0
\\ \nonumber
{\bf s}_{total} &=& \sum_{r = 1}^6 {\bf s}(r)
\eea
Obviously, the ``total angular momentum''
${\bf s}_{total}$ is invariant with respect to
any permutation of the six variables appearing
as indices in (98), i.e. we may simultaneously
characterize the space (89) the 
representations of the permutation group $S_6$
which acts on the states (98) by
\be
d(\nu)^{j_1 j_2 j_3 j_4 j_4 j_6}
\rightarrow d(\nu)^{j_{\sigma(1)} j_{\sigma(2)}
j_{\sigma(3)} j_{\sigma(4)} j_{\sigma(5)} j_{\sigma(6)}}
\ee
($\sigma \in S_6$). One may use the
decomposition of this representation
for classifying the states in the space (89). In its
general form, however, the corresponding permutation
operators do not commute with the Casimir operators
(47) since they involve only ``reduced total angular momenta''.
In our case we have e.g.
\bea
A_1 &=& ({\bf s}(2) +{\bf s}(3))^2
\\ \nonumber
A_2 & =& ({\bf s}(4) +{\bf s}(5))^2
\\ \nonumber
A_3 & =& ({\bf s}(2) +{\bf s}(3) + {\bf s}(4) +{\bf s}(5))^2
    - A_1 - A_2
\eea

Conveniently chosen subgroups of $S_6$, however,
do commute. We may take, e.g., $S_2$ embedded
in $S_6$ in different ways: If $ \pi$ is the
non-trivial element of $S_2$, we may put

$\pi (1,2,3,4,5,6) = (1,3,2,4,5,6)$ -
yielding $\chi^{3,4,1}$ and $\chi^{3,4,2}$ as
antisymmetric and  $ \chi^{3,4,3}$ and $ \chi^{3,4,4}$ 
as symmetric representations -

or $\pi (1,2,3,4,5,6) = (1,3,2,5,4,6)$ -
yielding $\chi^{3,4,1}$ and $\chi^{3,4,3}$
as antisymmetric and $\chi^{3,4,2}$ and $\chi^{3,4,4}$
as symmetric representations.

Of course, for SU(2), this does not yield independent
quantum numbers. With a suitable choice of the
permutation subgroup, however,
this may be the case for SU(3).

\subsubsection*{A4. Examples for the incorporation of the
Euclidean group}
We restrict ourselves to $SU(2)$ and $D = 2$.
A convenient enumeration of the Euclidean group for $D = 2$
is given by
\bea
  [n,\mu,m,\sigma] =    R^n t^\mu R^m P^\sigma \;\;\; ; \; m,n &=& 0,1,2,3
\\ \nonumber
      \sigma &=& 0,1
\\ \nonumber
       \mu &=& 0,1,2,3,4,5,.....
\eea
where the use the following conventions:

The {\em parity transformation} $P$  is 
fixed  by the condition
 $T(P) \Lambda^{3,4,2} = \Lambda^{3,4,3}$

{\em The rotation} $R$ is given by the constraint
that it has rotation angle $\pi/2$ and fulfils 
$ T(R)\chi^1 = \chi^1 = \Lambda^{1,1,1}$

The ``one-unit'' translation $t$ is defined by the
condition that $ \Lambda_G^{2,2} = \chi^1 \; T(t)\chi^1$.

Taking into account $ T(P) \chi^1 = \chi^1 $, we obtain in lowest 
order $\gamma_1 = \gamma_2 =(1,1,1)$ the non-trivially connected
cases ($n,m,\sigma$ are arbitrary with the restriction (96))
\bea
\gamma_3 &=& (2,1,1) \;\; u = [n,0,m,\sigma] \;\; v = [0,0,0,0]
 \\ \nonumber
\gamma_3 &=& (2,2,1) \;\; u = [n,1,m,\sigma] \;\; v = [4-n,0,0,0]
\eea

Combining first and second order on the r.h.s. of equation (69)
we have the following generic alternatives:
\\ 
$\gamma_1 = (2,1,1) ; \gamma_2 = (1,1,1)$:
\bea
\gamma_3 = (3,1,1) \;\; u &=& [n,0,m,\sigma] \;\; v = [0,0,0,0]
\\ \nonumber
\gamma_3 = (3,2,1) \;\; u &=& [n,1,m,\sigma] \;\; v = [4-n,0,0,0]
\eea
\\
$\gamma_1 = (2,2,1) ; \gamma_2 = (1,1,1)$:
\bea
\gamma_3 = (3,2,1) \;\; u &=& [0,0,m,\sigma] \;\; v = [0,0,0,0]
\\ \nonumber
 u &=& [0,1,m,\sigma] \;\; v = [0,1,2,0]
\\ \nonumber
\gamma_3 = (3,3,1) \;\; u &=& [0,2,m,\sigma] \;\; v = [0,0,0,0]
\\ \nonumber
 u &=& [2,0,m,\sigma] \;\; v = [0,1,2,0]
\\ \nonumber
\gamma_3 = (3,4,1) \;\; u &=& [1,1,m,\sigma] \;\; v = [0,0,0,0]
\\ \nonumber
 u &=& [3,1,m,\sigma] \;\; v = [0,0,1,1]
\eea
Here, the Euclidean elements $v$ are chosen such that a
non-trivial connection with {\em the same}
pairs $(u,v)$ is described for the $\gamma$-tripletts

(2,1,2),(1,1,1),[(3,1,2) or (3,2,2)] for the two cases (103) and

(2,2,2),(1,1,1),[(3,3,2) or (3,4,2)] for the  cases (104).

The Euclidean elements $(u,v)$ for the cases where
 $\gamma_1$ and $\gamma_2$ are exchanged may be
obtained by the replacements

$u \rightarrow u^{-1}, v \rightarrow u^{-1} v $

in the above formulas. This follows from
$\Lambda^2 T(u^{-1}) \Lambda^1 = T(u^{-1}) (\Lambda^1 T(u)\Lambda^2)
=  T(u^{-1} v)\Lambda^3$ if 
$\Lambda^1 T(u) \Lambda^2 = T(v)\Lambda^3$.

This provides all non-vanishing $\eta$-coefficients (66)
up to the order $\delta_3 = 3$.

\newpage

\begin{center}
  {\large {\bf FIGURES}}\\
\end{center}

\vspace{1.0cm}

\begin{center}
\setlength{\unitlength}{0.00083300in}%
\setlength{\unitlength}{0.0006in}

\begingroup\makeatletter\ifx\SetFigFont\undefined
\def\x#1#2#3#4#5#6#7\relax{\def\x{#1#2#3#4#5#6}}%
\expandafter\x\fmtname xxxxxx\relax \def\y{splain}%
\ifx\x\y   
\gdef\SetFigFont#1#2#3{%
  \ifnum #1<17\tiny\else \ifnum #1<20\small\else
  \ifnum #1<24\normalsize\else \ifnum #1<29\large\else
  \ifnum #1<34\Large\else \ifnum #1<41\LARGE\else
     \huge\fi\fi\fi\fi\fi\fi
  \csname #3\endcsname}%
\else
\gdef\SetFigFont#1#2#3{\begingroup
  \count@#1\relax \ifnum 25<\count@\count@25\fi
  \def\x{\endgroup\@setsize\SetFigFont{#2pt}}%
  \expandafter\x
    \csname \romannumeral\the\count@ pt\expandafter\endcsname
    \csname @\romannumeral\the\count@ pt\endcsname
  \csname #3\endcsname}%
\fi
\fi\endgroup
\begin{picture}(5175,3240)(2926,-5869)
\thicklines
\put(3301,-5461){\vector( 0, 1){1200}}
\put(5401,-4261){\line( 0, 1){1200}}
\put(5401,-5461){\vector( 0, 1){1200}}
\put(4276,-5461){\line( 1, 0){1200}}
\put(5476,-5461){\vector( 1, 0){1200}}
\put(6601,-5461){\line( 1, 0){1200}}
\put(7801,-3061){\line( 0,-1){1200}}
\put(7801,-5461){\vector( 0, 1){1200}}
\put(3301,-5461){\vector( 1, 0){1200}}
\put(3301,-3061){\vector( 1, 0){1200}}
\put(3301,-4261){\line( 0, 1){1200}}
\put(5476,-3061){\vector( 1, 0){1200}}
\put(6601,-3061){\line( 1, 0){1200}}
\put(4426,-3061){\line( 1, 0){1200}}
\put(8000,-4340){\makebox(0,0)[lb]{\smash{\SetFigFont{14}{16.8}{rm}5}}}
\put(2926,-4340){\makebox(0,0)[lb]{\smash{\SetFigFont{14}{16.8}{rm}2}}}
\put(5601,-4340){\makebox(0,0)[lb]{\smash{\SetFigFont{14}{16.8}{rm}7}}}
\put(4240,-5836){\makebox(0,0)[lb]{\smash{\SetFigFont{14}{16.8}{rm}3}}}
\put(6450,-5836){\makebox(0,0)[lb]{\smash{\SetFigFont{14}{16.8}{rm}4}}}
\put(4276,-2836){\makebox(0,0)[lb]{\smash{\SetFigFont{12}{14.4}{rm}1}}}
\put(6451,-2836){\makebox(0,0)[lb]{\smash{\SetFigFont{14}{16.8}{rm}6}}}
\end{picture}

\end{center}

Figure 1.
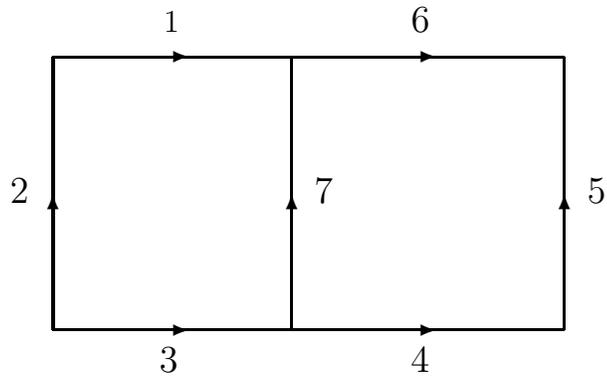
Link pattern of the D-loop functions (41).
The numbers indicate the enumeration of the link
 angular momenta $(J_1,..,J_7)$.

\newpage
\begin{tabbing}
333\quad \= 333\quad \=  
11111111111111111 \quad \=
1111111111111111111 \quad \=
1111111111111111111 \quad \=
1111111111111111111 \quad \=
1111111111111111111 \quad \=
\kill

 \> \> \> $k$ \\  
$\delta$ \> \> 1 \> 2 \> 3\> 4\> \\  \\ \\ \\
1 \> \setlength{\unitlength}{0.00083300in}%
\setlength{\unitlength}{0.000250in}%
\begingroup\makeatletter\ifx\SetFigFont\undefined
\def\x#1#2#3#4#5#6#7\relax{\def\x{#1#2#3#4#5#6}}%
\expandafter\x\fmtname xxxxxx\relax \def\y{splain}%
\ifx\x\y   
\gdef\SetFigFont#1#2#3{%
  \ifnum #1<17\tiny\else \ifnum #1<20\small\else
  \ifnum #1<24\normalsize\else \ifnum #1<29\large\else
  \ifnum #1<34\Large\else \ifnum #1<41\LARGE\else
     \huge\fi\fi\fi\fi\fi\fi
  \csname #3\endcsname}%
\else
\gdef\SetFigFont#1#2#3{\begingroup
  \count@#1\relax \ifnum 25<\count@\count@25\fi
  \def\x{\endgroup\@setsize\SetFigFont{#2pt}}%
  \expandafter\x
    \csname \romannumeral\the\count@ pt\expandafter\endcsname
    \csname @\romannumeral\the\count@ pt\endcsname
  \csname #3\endcsname}%
\fi
\fi\endgroup
\begin{picture}(1444,1444)(3579,-4883)
\thicklines
\put(3601,-4861){\vector( 1, 0){1200}}
\put(6001,-4861){\vector( 0, 1){1200}}
\put(6001,-3661){\line( 0, 1){1200}}
\put(3601,-2461){\vector( 0,-1){1200}}
\put(3601,-4861){\line( 0, 1){1200}}
\put(6001,-2461){\vector(-1, 0){1200}}
\put(4801,-4861){\line( 1, 0){1200}}
\put(3601,-2461){\line( 1, 0){1200}}
\end{picture}
 \> \\ \\ \\ \\ 
2 \> \setlength{\unitlength}{0.00083300in}%
\setlength{\unitlength}{0.000250in}%
\begingroup\makeatletter\ifx\SetFigFont\undefined
\def\x#1#2#3#4#5#6#7\relax{\def\x{#1#2#3#4#5#6}}%
\expandafter\x\fmtname xxxxxx\relax \def\y{splain}%
\ifx\x\y   
\gdef\SetFigFont#1#2#3{%
  \ifnum #1<17\tiny\else \ifnum #1<20\small\else
  \ifnum #1<24\normalsize\else \ifnum #1<29\large\else
  \ifnum #1<34\Large\else \ifnum #1<41\LARGE\else
     \huge\fi\fi\fi\fi\fi\fi
  \csname #3\endcsname}%
\else
\gdef\SetFigFont#1#2#3{\begingroup
  \count@#1\relax \ifnum 25<\count@\count@25\fi
  \def\x{\endgroup\@setsize\SetFigFont{#2pt}}%
  \expandafter\x
    \csname \romannumeral\the\count@ pt\expandafter\endcsname
    \csname @\romannumeral\the\count@ pt\endcsname
  \csname #3\endcsname}%
\fi
\fi\endgroup
\begin{picture}(1444,1444)(3579,-4883)
\thicklines
\put(3601,-4861){\vector( 1, 0){1200}}
\put(6001,-4861){\vector( 0, 1){1200}}
\put(6001,-3661){\line( 0, 1){1200}}
\put(3601,-2461){\vector( 0,-1){1200}}
\put(3601,-4861){\line( 0, 1){1200}}
\put(6001,-2461){\vector(-1, 0){1200}}
\put(4801,-4861){\line( 1, 0){1200}}
\put(3601,-2461){\line( 1, 0){1200}}
\put(3901,-2761){\line( 1, 0){900}}
\put(4801,-2761){\line( 0, 1){  0}}
\put(5701,-2761){\vector(-1, 0){900}}
\put(5701,-2761){\line( 0,-1){900}}
\put(5701,-4561){\vector( 0, 1){900}}
\put(3901,-2761){\vector( 0,-1){900}}
\put(3901,-3661){\line( 0,-1){900}}
\put(3901,-4561){\vector( 1, 0){900}}
\put(4801,-4561){\line( 1, 0){900}}
\put(5701,-4561){\line( 0, 1){  0}}
\end{picture}
 \> \hskip 2cm \setlength{\unitlength}{0.00083300in}%
\setlength{\unitlength}{0.000250in}%
\begingroup\makeatletter\ifx\SetFigFont\undefined
\def\x#1#2#3#4#5#6#7\relax{\def\x{#1#2#3#4#5#6}}%
\expandafter\x\fmtname xxxxxx\relax \def\y{splain}%
\ifx\x\y   
\gdef\SetFigFont#1#2#3{%
  \ifnum #1<17\tiny\else \ifnum #1<20\small\else
  \ifnum #1<24\normalsize\else \ifnum #1<29\large\else
  \ifnum #1<34\Large\else \ifnum #1<41\LARGE\else
     \huge\fi\fi\fi\fi\fi\fi
  \csname #3\endcsname}%
\else
\gdef\SetFigFont#1#2#3{\begingroup
  \count@#1\relax \ifnum 25<\count@\count@25\fi
  \def\x{\endgroup\@setsize\SetFigFont{#2pt}}%
  \expandafter\x
    \csname \romannumeral\the\count@ pt\expandafter\endcsname
    \csname @\romannumeral\the\count@ pt\endcsname
  \csname #3\endcsname}%
\fi
\fi\endgroup
\begin{picture}(1144,1444)(3579,-4883)
\thicklines
\put(3601,-4861){\vector( 1, 0){1200}}
\put(6001,-4861){\vector( 0, 1){1200}}
\put(3601,-4861){\line( 0, 1){1200}}
\put(4801,-4861){\line( 1, 0){1200}}
\put(6001,-2461){\vector(-1, 0){1200}}
\put(3601,-2461){\line( 1, 0){1200}}
\put(3601,-2461){\vector( 0,-1){1200}}
\put(6001,-3661){\line( 0, 1){1200}}
\put(6301,-2461){\vector( 0,-1){1200}}
\put(6301,-4861){\line( 0, 1){1200}}
\put(6301,-4861){\vector( 1, 0){1200}}
\put(7501,-4861){\line( 1, 0){1200}}
\put(6301,-2461){\line( 1, 0){1200}}
\put(8701,-2461){\vector(-1, 0){1200}}
\put(8701,-3661){\line( 0, 1){1200}}
\put(8701,-4861){\vector( 0, 1){1200}}
\end{picture}
 \> \\ \\  
3 \>  \setlength{\unitlength}{0.00083300in}%
\setlength{\unitlength}{0.0002500in}%
\begingroup\makeatletter\ifx\SetFigFont\undefined
\def\x#1#2#3#4#5#6#7\relax{\def\x{#1#2#3#4#5#6}}%
\expandafter\x\fmtname xxxxxx\relax \def\y{splain}%
\ifx\x\y   
\gdef\SetFigFont#1#2#3{%
  \ifnum #1<17\tiny\else \ifnum #1<20\small\else
  \ifnum #1<24\normalsize\else \ifnum #1<29\large\else
  \ifnum #1<34\Large\else \ifnum #1<41\LARGE\else
     \huge\fi\fi\fi\fi\fi\fi
  \csname #3\endcsname}%
\else
\gdef\SetFigFont#1#2#3{\begingroup
  \count@#1\relax \ifnum 25<\count@\count@25\fi
  \def\x{\endgroup\@setsize\SetFigFont{#2pt}}%
  \expandafter\x
    \csname \romannumeral\the\count@ pt\expandafter\endcsname
    \csname @\romannumeral\the\count@ pt\endcsname
  \csname #3\endcsname}%
\fi
\fi\endgroup
\begin{picture}(2444,2444)(3579,-4883)
\thicklines
\put(3601,-4861){\vector( 1, 0){1200}}
\put(6001,-4861){\vector( 0, 1){1200}}
\put(6001,-3661){\line( 0, 1){1200}}
\put(3601,-2461){\vector( 0,-1){1200}}
\put(3601,-4861){\line( 0, 1){1200}}
\put(6001,-2461){\vector(-1, 0){1200}}
\put(4801,-4861){\line( 1, 0){1200}}
\put(3601,-2461){\line( 1, 0){1200}}
\put(3901,-2761){\line( 1, 0){900}}
\put(4801,-2761){\line( 0, 1){  0}}
\put(5701,-2761){\vector(-1, 0){900}}
\put(5701,-2761){\line( 0,-1){900}}
\put(5701,-4561){\vector( 0, 1){900}}
\put(3901,-2761){\vector( 0,-1){900}}
\put(3901,-3661){\line( 0,-1){900}}
\put(3901,-4561){\vector( 1, 0){900}}
\put(4801,-4561){\line( 1, 0){900}}
\put(5701,-4561){\line( 0, 1){  0}}
\put(5851,-2611){\vector(-1, 0){1050}}
\put(5851,-3661){\line( 0, 1){1050}}
\put(5851,-4711){\vector( 0, 1){1050}}
\put(3751,-2611){\line( 1, 0){1050}}
\put(3751,-2611){\vector( 0,-1){1050}}
\put(3751,-3511){\line( 0,-1){1200}}
\put(3751,-4711){\vector( 1, 0){1050}}
\put(4801,-4711){\line( 1, 0){1050}}
\end{picture}
 \> \hskip 2cm\setlength{\unitlength}{0.00083300in}%
\setlength{\unitlength}{0.0002500in}%
\begingroup\makeatletter\ifx\SetFigFont\undefined
\def\x#1#2#3#4#5#6#7\relax{\def\x{#1#2#3#4#5#6}}%
\expandafter\x\fmtname xxxxxx\relax \def\y{splain}%
\ifx\x\y   
\gdef\SetFigFont#1#2#3{%
  \ifnum #1<17\tiny\else \ifnum #1<20\small\else
  \ifnum #1<24\normalsize\else \ifnum #1<29\large\else
  \ifnum #1<34\Large\else \ifnum #1<41\LARGE\else
     \huge\fi\fi\fi\fi\fi\fi
  \csname #3\endcsname}%
\else
\gdef\SetFigFont#1#2#3{\begingroup
  \count@#1\relax \ifnum 25<\count@\count@25\fi
  \def\x{\endgroup\@setsize\SetFigFont{#2pt}}%
  \expandafter\x
    \csname \romannumeral\the\count@ pt\expandafter\endcsname
    \csname @\romannumeral\the\count@ pt\endcsname
  \csname #3\endcsname}%
\fi
\fi\endgroup
\begin{picture}(4844,2444)(3579,-4883)
\thicklines
\put(3601,-4861){\vector( 1, 0){1200}}
\put(6001,-4861){\vector( 0, 1){1200}}
\put(6001,-3661){\line( 0, 1){1200}}
\put(3601,-2461){\vector( 0,-1){1200}}
\put(3601,-4861){\line( 0, 1){1200}}
\put(6001,-2461){\vector(-1, 0){1200}}
\put(4801,-4861){\line( 1, 0){1200}}
\put(3601,-2461){\line( 1, 0){1200}}
\put(3901,-2761){\line( 1, 0){900}}
\put(4801,-2761){\line( 0, 1){  0}}
\put(5701,-2761){\vector(-1, 0){900}}
\put(5701,-2761){\line( 0,-1){900}}
\put(5701,-4561){\vector( 0, 1){900}}
\put(3901,-2761){\vector( 0,-1){900}}
\put(3901,-3661){\line( 0,-1){900}}
\put(3901,-4561){\vector( 1, 0){900}}
\put(4801,-4561){\line( 1, 0){900}}
\put(5701,-4561){\line( 0, 1){  0}}
\put(6301,-2611){\vector( 0,-1){1050}}
\put(6301,-3511){\line( 0,-1){1200}}
\put(8401,-3661){\line( 0, 1){1050}}
\put(8401,-4711){\vector( 0, 1){1050}}
\put(7351,-4711){\line( 1, 0){1050}}
\put(8401,-2611){\vector(-1, 0){1050}}
\put(6301,-2611){\line( 1, 0){1050}}
\put(6301,-4711){\vector( 1, 0){1050}}
\end{picture}
 \> \hskip 2cm
\setlength{\unitlength}{0.00083300in}%
\setlength{\unitlength}{0.0002500in}%
\begingroup\makeatletter\ifx\SetFigFont\undefined
\def\x#1#2#3#4#5#6#7\relax{\def\x{#1#2#3#4#5#6}}%
\expandafter\x\fmtname xxxxxx\relax \def\y{splain}%
\ifx\x\y   
\gdef\SetFigFont#1#2#3{%
  \ifnum #1<17\tiny\else \ifnum #1<20\small\else
  \ifnum #1<24\normalsize\else \ifnum #1<29\large\else
  \ifnum #1<34\Large\else \ifnum #1<41\LARGE\else
     \huge\fi\fi\fi\fi\fi\fi
  \csname #3\endcsname}%
\else
\gdef\SetFigFont#1#2#3{\begingroup
  \count@#1\relax \ifnum 25<\count@\count@25\fi
  \def\x{\endgroup\@setsize\SetFigFont{#2pt}}%
  \expandafter\x
    \csname \romannumeral\the\count@ pt\expandafter\endcsname
    \csname @\romannumeral\the\count@ pt\endcsname
  \csname #3\endcsname}%
\fi
\fi\endgroup
\begin{picture}(7844,2444)(3579,-4883)
\thicklines
\put(3601,-4861){\vector( 1, 0){1200}}
\put(6001,-4861){\vector( 0, 1){1200}}
\put(3601,-4861){\line( 0, 1){1200}}
\put(4801,-4861){\line( 1, 0){1200}}
\put(6001,-2461){\vector(-1, 0){1200}}
\put(3601,-2461){\line( 1, 0){1200}}
\put(3601,-2461){\vector( 0,-1){1200}}
\put(6001,-3661){\line( 0, 1){1200}}
\put(6301,-2461){\vector( 0,-1){1200}}
\put(6301,-4861){\line( 0, 1){1200}}
\put(6301,-4861){\vector( 1, 0){1200}}
\put(7501,-4861){\line( 1, 0){1200}}
\put(6301,-2461){\line( 1, 0){1200}}
\put(8701,-2461){\vector(-1, 0){1200}}
\put(8701,-3661){\line( 0, 1){1200}}
\put(8701,-4861){\vector( 0, 1){1200}}
\put(9001,-2461){\vector( 0,-1){1200}}
\put(9001,-4861){\line( 0, 1){1200}}
\put(9001,-4861){\vector( 1, 0){1200}}
\put(10201,-4861){\line( 1, 0){1200}}
\put(9001,-2461){\line( 1, 0){1200}}
\put(11401,-2461){\vector(-1, 0){1200}}
\put(11401,-3661){\line( 0, 1){1200}}
\put(11401,-4861){\vector( 0, 1){1200}}
\end{picture}
 \> \hskip 3.5cm \setlength{\unitlength}{0.00083300in}%
\setlength{\unitlength}{0.0002500in}%
\begingroup\makeatletter\ifx\SetFigFont\undefined
\def\x#1#2#3#4#5#6#7\relax{\def\x{#1#2#3#4#5#6}}%
\expandafter\x\fmtname xxxxxx\relax \def\y{splain}%
\ifx\x\y   
\gdef\SetFigFont#1#2#3{%
  \ifnum #1<17\tiny\else \ifnum #1<20\small\else
  \ifnum #1<24\normalsize\else \ifnum #1<29\large\else
  \ifnum #1<34\Large\else \ifnum #1<41\LARGE\else
     \huge\fi\fi\fi\fi\fi\fi
  \csname #3\endcsname}%
\else
\gdef\SetFigFont#1#2#3{\begingroup
  \count@#1\relax \ifnum 25<\count@\count@25\fi
  \def\x{\endgroup\@setsize\SetFigFont{#2pt}}%
  \expandafter\x
    \csname \romannumeral\the\count@ pt\expandafter\endcsname
    \csname @\romannumeral\the\count@ pt\endcsname
  \csname #3\endcsname}%
\fi
\fi\endgroup
\begin{picture}(5144,5144)(3579,-4883)
\thicklines
\put(3601,-4861){\vector( 1, 0){1200}}
\put(6001,-4861){\vector( 0, 1){1200}}
\put(3601,-4861){\line( 0, 1){1200}}
\put(4801,-4861){\line( 1, 0){1200}}
\put(6001,-2461){\vector(-1, 0){1200}}
\put(3601,-2461){\line( 1, 0){1200}}
\put(3601,-2461){\vector( 0,-1){1200}}
\put(6001,-3661){\line( 0, 1){1200}}
\put(6301,-2461){\vector( 0,-1){1200}}
\put(6301,-4861){\line( 0, 1){1200}}
\put(6301,-4861){\vector( 1, 0){1200}}
\put(7501,-4861){\line( 1, 0){1200}}
\put(6301,-2461){\line( 1, 0){1200}}
\put(8701,-2461){\vector(-1, 0){1200}}
\put(8701,-3661){\line( 0, 1){1200}}
\put(8701,-4861){\vector( 0, 1){1200}}
\put(3601,239){\vector( 0,-1){1200}}
\put(3601,-2161){\line( 0, 1){1200}}
\put(3601,-2161){\vector( 1, 0){1200}}
\put(4801,-2161){\line( 1, 0){1200}}
\put(6001,-2161){\vector( 0, 1){1200}}
\put(6001,-961){\line( 0, 1){1200}}
\put(3601,239){\line( 1, 0){1200}}
\put(6001,239){\vector(-1, 0){1200}}
\end{picture}
 \>  \\ \\  

\end{tabbing}
Figure 2.
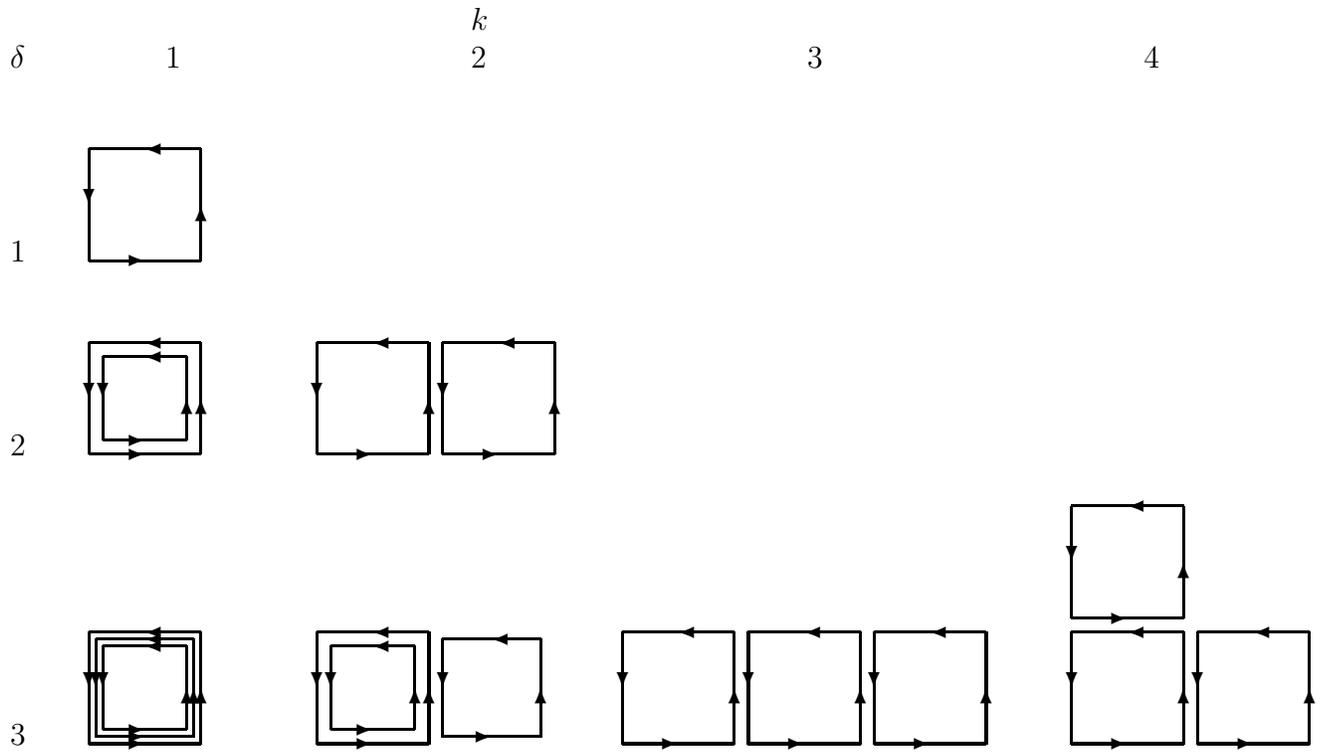
Loop structure of the generic character functions
$\Lambda_G^{\delta,k}$ (see equation (53))
up to order $\delta = 3$ for SU(2).

\newpage
\begin{tabbing}
333\quad \= 444\quad \= 11111111111111111111\quad \=
(0,33,33)\quad \= (0,33,33)\quad \=
(0,0,0)\quad \=(0,0,0)\quad \= \kill
 \>  \> \> \> $\nu$ \>\>\>  \\ 
$\delta$ \>k  \>  \> 1 \> 2 \> 3 \> 4 \>  dependent \\ \\ \\
1 \> 1 \> \setlength{\unitlength}{0.00083300in}%
\setlength{\unitlength}{0.000250in}%
\begingroup\makeatletter\ifx\SetFigFont\undefined
\def\x#1#2#3#4#5#6#7\relax{\def\x{#1#2#3#4#5#6}}%
\expandafter\x\fmtname xxxxxx\relax \def\y{splain}%
\ifx\x\y   
\gdef\SetFigFont#1#2#3{%
  \ifnum #1<17\tiny\else \ifnum #1<20\small\else
  \ifnum #1<24\normalsize\else \ifnum #1<29\large\else
  \ifnum #1<34\Large\else \ifnum #1<41\LARGE\else
     \huge\fi\fi\fi\fi\fi\fi
  \csname #3\endcsname}%
\else
\gdef\SetFigFont#1#2#3{\begingroup
  \count@#1\relax \ifnum 25<\count@\count@25\fi
  \def\x{\endgroup\@setsize\SetFigFont{#2pt}}%
  \expandafter\x
    \csname \romannumeral\the\count@ pt\expandafter\endcsname
    \csname @\romannumeral\the\count@ pt\endcsname
  \csname #3\endcsname}%
\fi
\fi\endgroup
\begin{picture}(1544,1544)(2979,-3383)
\thicklines
\put(3001,-3361){\framebox(1500,1500){}}
\end{picture}
 \> $(\frac{1}{2}^4) = \chi^1$ \\ \\ \\  
2 \> 1  \> \setlength{\unitlength}{0.00083300in}%
\setlength{\unitlength}{0.000250in}%
\begingroup\makeatletter\ifx\SetFigFont\undefined
\def\x#1#2#3#4#5#6#7\relax{\def\x{#1#2#3#4#5#6}}%
\expandafter\x\fmtname xxxxxx\relax \def\y{splain}%
\ifx\x\y   
\gdef\SetFigFont#1#2#3{%
  \ifnum #1<17\tiny\else \ifnum #1<20\small\else
  \ifnum #1<24\normalsize\else \ifnum #1<29\large\else
  \ifnum #1<34\Large\else \ifnum #1<41\LARGE\else
     \huge\fi\fi\fi\fi\fi\fi
  \csname #3\endcsname}%
\else
\gdef\SetFigFont#1#2#3{\begingroup
  \count@#1\relax \ifnum 25<\count@\count@25\fi
  \def\x{\endgroup\@setsize\SetFigFont{#2pt}}%
  \expandafter\x
    \csname \romannumeral\the\count@ pt\expandafter\endcsname
    \csname @\romannumeral\the\count@ pt\endcsname
  \csname #3\endcsname}%
\fi
\fi\endgroup
\begin{picture}(1544,1544)(2979,-3383)
\thicklines
\put(3001,-3361){\framebox(1500,1500){}}
\end{picture}
 \> ($1^4$)\>\>\>\> $(0^4) = \chi^0$\\ \\ \\ 
  \> 2  \> \setlength{\unitlength}{0.00083300in}%
\setlength{\unitlength}{0.0002500in}%
\begingroup\makeatletter\ifx\SetFigFont\undefined
\def\x#1#2#3#4#5#6#7\relax{\def\x{#1#2#3#4#5#6}}%
\expandafter\x\fmtname xxxxxx\relax \def\y{splain}%
\ifx\x\y   
\gdef\SetFigFont#1#2#3{%
  \ifnum #1<17\tiny\else \ifnum #1<20\small\else
  \ifnum #1<24\normalsize\else \ifnum #1<29\large\else
  \ifnum #1<34\Large\else \ifnum #1<41\LARGE\else
     \huge\fi\fi\fi\fi\fi\fi
  \csname #3\endcsname}%
\else
\gdef\SetFigFont#1#2#3{\begingroup
  \count@#1\relax \ifnum 25<\count@\count@25\fi
  \def\x{\endgroup\@setsize\SetFigFont{#2pt}}%
  \expandafter\x
    \csname \romannumeral\the\count@ pt\expandafter\endcsname
    \csname @\romannumeral\the\count@ pt\endcsname
  \csname #3\endcsname}%
\fi
\fi\endgroup
\begin{picture}(3044,1544)(2979,-3383)
\thicklines
\put(3001,-3361){\framebox(1500,1500){}}
\put(4501,-3361){\framebox(1500,1500){}}
\put(4156,-2686){\makebox(0,0)[lb]{\smash{\SetFigFont{12}{16.8}{rm}1}}}
\end{picture}
 \> $(0,\frac{1}{2}^3,\frac{1}{2}^3)$\> 
$(1,\frac{1}{2}^3,\frac{1}{2}^3)$\\  \\ \\ 
3 \> 1 \> \setlength{\unitlength}{0.00083300in}%
\setlength{\unitlength}{0.000250in}%
\begingroup\makeatletter\ifx\SetFigFont\undefined
\def\x#1#2#3#4#5#6#7\relax{\def\x{#1#2#3#4#5#6}}%
\expandafter\x\fmtname xxxxxx\relax \def\y{splain}%
\ifx\x\y   
\gdef\SetFigFont#1#2#3{%
  \ifnum #1<17\tiny\else \ifnum #1<20\small\else
  \ifnum #1<24\normalsize\else \ifnum #1<29\large\else
  \ifnum #1<34\Large\else \ifnum #1<41\LARGE\else
     \huge\fi\fi\fi\fi\fi\fi
  \csname #3\endcsname}%
\else
\gdef\SetFigFont#1#2#3{\begingroup
  \count@#1\relax \ifnum 25<\count@\count@25\fi
  \def\x{\endgroup\@setsize\SetFigFont{#2pt}}%
  \expandafter\x
    \csname \romannumeral\the\count@ pt\expandafter\endcsname
    \csname @\romannumeral\the\count@ pt\endcsname
  \csname #3\endcsname}%
\fi
\fi\endgroup
\begin{picture}(1544,1544)(2979,-3383)
\thicklines
\put(3001,-3361){\framebox(1500,1500){}}
\end{picture}
  \> $(\frac{3}{2}^4)$\>\>\>\>
$(\frac{1}{2}^4)$\\ \\ \\  
\>   2 \>\setlength{\unitlength}{0.00083300in}%
\setlength{\unitlength}{0.0002500in}%
\begingroup\makeatletter\ifx\SetFigFont\undefined
\def\x#1#2#3#4#5#6#7\relax{\def\x{#1#2#3#4#5#6}}%
\expandafter\x\fmtname xxxxxx\relax \def\y{splain}%
\ifx\x\y   
\gdef\SetFigFont#1#2#3{%
  \ifnum #1<17\tiny\else \ifnum #1<20\small\else
  \ifnum #1<24\normalsize\else \ifnum #1<29\large\else
  \ifnum #1<34\Large\else \ifnum #1<41\LARGE\else
     \huge\fi\fi\fi\fi\fi\fi
  \csname #3\endcsname}%
\else
\gdef\SetFigFont#1#2#3{\begingroup
  \count@#1\relax \ifnum 25<\count@\count@25\fi
  \def\x{\endgroup\@setsize\SetFigFont{#2pt}}%
  \expandafter\x
    \csname \romannumeral\the\count@ pt\expandafter\endcsname
    \csname @\romannumeral\the\count@ pt\endcsname
  \csname #3\endcsname}%
\fi
\fi\endgroup
\begin{picture}(3322,1544)(2701,-3383)
\thicklines
\put(3001,-3361){\framebox(1500,1500){}}
\put(4501,-3361){\framebox(1500,1500){}}
\put(4156,-2800){\makebox(0,0)[lb]{\smash{\SetFigFont{12}{16.8}{rm}1}}}
\put(2551,-2800){\makebox(0,0)[lb]{\smash{\SetFigFont{12}{16.8}{rm}2}}}
\end{picture}
  \>  $(\frac{1}{2},1^3,\frac{1}{2}^3)$\>
$(\frac{3}{2},1^3,\frac{1}{2}^3)$\>\>\> 
$(\frac{1}{2},0^3,\frac{1}{2}^3) = (\frac{1}{2}^4) $\\ \\ \\ 
\> 3 \>\setlength{\unitlength}{0.00083300in}%
\setlength{\unitlength}{0.0002500in}%
\begingroup\makeatletter\ifx\SetFigFont\undefined
\def\x#1#2#3#4#5#6#7\relax{\def\x{#1#2#3#4#5#6}}%
\expandafter\x\fmtname xxxxxx\relax \def\y{splain}%
\ifx\x\y   
\gdef\SetFigFont#1#2#3{%
  \ifnum #1<17\tiny\else \ifnum #1<20\small\else
  \ifnum #1<24\normalsize\else \ifnum #1<29\large\else
  \ifnum #1<34\Large\else \ifnum #1<41\LARGE\else
     \huge\fi\fi\fi\fi\fi\fi
  \csname #3\endcsname}%
\else
\gdef\SetFigFont#1#2#3{\begingroup
  \count@#1\relax \ifnum 25<\count@\count@25\fi
  \def\x{\endgroup\@setsize\SetFigFont{#2pt}}%
  \expandafter\x
    \csname \romannumeral\the\count@ pt\expandafter\endcsname
    \csname @\romannumeral\the\count@ pt\endcsname
  \csname #3\endcsname}%
\fi
\fi\endgroup
\begin{picture}(4544,1544)(2979,-3383)
\thicklines
\put(3001,-3361){\framebox(1500,1500){}}
\put(4501,-3361){\framebox(1500,1500){}}
\put(6001,-3361){\framebox(1500,1500){}}
\put(4076,-2800){\makebox(0,0)[lb]{\smash{\SetFigFont{12}{16.8}{rm}1}}}
\put(5606,-2800){\makebox(0,0)[lb]{\smash{\SetFigFont{12}{16.8}{rm}2}}}
\end{picture}
 \> (0,0)\> (0,1) \> (1,1) \\ \\ \\ 
\> 4 \>\setlength{\unitlength}{0.00083300in}%
\setlength{\unitlength}{0.0002500in}%
\begingroup\makeatletter\ifx\SetFigFont\undefined
\def\x#1#2#3#4#5#6#7\relax{\def\x{#1#2#3#4#5#6}}%
\expandafter\x\fmtname xxxxxx\relax \def\y{splain}%
\ifx\x\y   
\gdef\SetFigFont#1#2#3{%
  \ifnum #1<17\tiny\else \ifnum #1<20\small\else
  \ifnum #1<24\normalsize\else \ifnum #1<29\large\else
  \ifnum #1<34\Large\else \ifnum #1<41\LARGE\else
     \huge\fi\fi\fi\fi\fi\fi
  \csname #3\endcsname}%
\else
\gdef\SetFigFont#1#2#3{\begingroup
  \count@#1\relax \ifnum 25<\count@\count@25\fi
  \def\x{\endgroup\@setsize\SetFigFont{#2pt}}%
  \expandafter\x
    \csname \romannumeral\the\count@ pt\expandafter\endcsname
    \csname @\romannumeral\the\count@ pt\endcsname
  \csname #3\endcsname}%
\fi
\fi\endgroup
\begin{picture}(3044,3044)(2979,-3383)
\thicklines
\put(3001,-3361){\framebox(1500,1500){}}
\put(4501,-3361){\framebox(1500,1500){}}
\put(3001,-1861){\framebox(1500,1500){}}
\put(3676,-1706){\makebox(0,0)[lb]{\smash{\SetFigFont{12}{16.8}{rm}1}}}
\put(4726,-2786){\makebox(0,0)[lb]{\smash{\SetFigFont{12}{16.8}{rm}2}}}
\end{picture}
  \> (0,0,0)\> (0,1,1)\> (1,1,0)\> (1,1,1)\\ \\ 
\end{tabbing}

Figure 3. 
$SU(2)$ D-loop basis functions $\chi^{\delta,k,\nu}$
characterized by the link
patterns and   and the related possible 
Casimir eigenvalue patterns which are generated up the order 
$\delta = 3$.
Non-equal angular momenta are enumerated in the  
link patterns and indicated in that order in the
eigenvalue patterns. Upper indices stand for the
degeneracy of these angular momenta.

For $\delta = 3, k = 3$ or $4$ all angular momenta
which are not specified are equal to $\frac{1}{2}$.

For $(\delta,k) = (3,4)$, the third angular momentum
is given by the ``intermediate''
coupling ${\bf j}_1 + {\bf j}_2$ (eq. (49)).

The last column gives the linear dependent D-loop
functions emerging up to this order. 

The orientation of the links is not marked, it may be
taken analogously to Fig. 1.

\newpage
\begin{tabbing}
33  \= 33  \= 3\quad \= 
111111111111111\quad \=
111111111111111\quad \=
111111111111111\quad \=
111111111111111\quad \=
\kill
 \> \> \> $\nu$ \\  
$\delta$ \> k  \> \> 1 \> 2 \> 3\> 4\> 5 \\  \\ 
2\> 1 \> \setlength{\unitlength}{0.00083300in}%
\setlength{\unitlength}{0.00020in}%
\begingroup\makeatletter\ifx\SetFigFont\undefined
\def\x#1#2#3#4#5#6#7\relax{\def\x{#1#2#3#4#5#6}}%
\expandafter\x\fmtname xxxxxx\relax \def\y{splain}%
\ifx\x\y   
\gdef\SetFigFont#1#2#3{%
  \ifnum #1<17\tiny\else \ifnum #1<20\small\else
  \ifnum #1<24\normalsize\else \ifnum #1<29\large\else
  \ifnum #1<34\Large\else \ifnum #1<41\LARGE\else
     \huge\fi\fi\fi\fi\fi\fi
  \csname #3\endcsname}%
\else
\gdef\SetFigFont#1#2#3{\begingroup
  \count@#1\relax \ifnum 25<\count@\count@25\fi
  \def\x{\endgroup\@setsize\SetFigFont{#2pt}}%
  \expandafter\x
    \csname \romannumeral\the\count@ pt\expandafter\endcsname
    \csname @\romannumeral\the\count@ pt\endcsname
  \csname #3\endcsname}%
\fi
\fi\endgroup
\begin{picture}(1444,1444)(3579,-4883)
\thicklines
\put(3601,-4861){\vector( 1, 0){1200}}
\put(6001,-4861){\vector( 0, 1){1200}}
\put(6001,-3661){\line( 0, 1){1200}}
\put(3601,-2461){\vector( 0,-1){1200}}
\put(3601,-4861){\line( 0, 1){1200}}
\put(6001,-2461){\vector(-1, 0){1200}}
\put(4801,-4861){\line( 1, 0){1200}}
\put(3601,-2461){\line( 1, 0){1200}}
\put(3901,-2761){\line( 1, 0){900}}
\put(4801,-2761){\line( 0, 1){  0}}
\put(5701,-2761){\vector(-1, 0){900}}
\put(5701,-2761){\line( 0,-1){900}}
\put(5701,-4561){\vector( 0, 1){900}}
\put(3901,-2761){\vector( 0,-1){900}}
\put(3901,-3661){\line( 0,-1){900}}
\put(3901,-4561){\vector( 1, 0){900}}
\put(4801,-4561){\line( 1, 0){900}}
\put(5701,-4561){\line( 0, 1){  0}}
\end{picture}
 \>\hskip 3cm \setlength{\unitlength}{0.00083300in}%
\setlength{\unitlength}{0.000200in}%
\begingroup\makeatletter\ifx\SetFigFont\undefined
\def\x#1#2#3#4#5#6#7\relax{\def\x{#1#2#3#4#5#6}}%
\expandafter\x\fmtname xxxxxx\relax \def\y{splain}%
\ifx\x\y   
\gdef\SetFigFont#1#2#3{%
  \ifnum #1<17\tiny\else \ifnum #1<20\small\else
  \ifnum #1<24\normalsize\else \ifnum #1<29\large\else
  \ifnum #1<34\Large\else \ifnum #1<41\LARGE\else
     \huge\fi\fi\fi\fi\fi\fi
  \csname #3\endcsname}%
\else
\gdef\SetFigFont#1#2#3{\begingroup
  \count@#1\relax \ifnum 25<\count@\count@25\fi
  \def\x{\endgroup\@setsize\SetFigFont{#2pt}}%
  \expandafter\x
    \csname \romannumeral\the\count@ pt\expandafter\endcsname
    \csname @\romannumeral\the\count@ pt\endcsname
  \csname #3\endcsname}%
\fi
\fi\endgroup
\begin{picture}(2444,2444)(3579,-4883)
\thicklines
\put(3601,-4861){\vector( 1, 0){1200}}
\put(6001,-4861){\vector( 0, 1){1200}}
\put(3601,-2461){\vector( 0,-1){1200}}
\put(3601,-4861){\line( 0, 1){1200}}
\put(4801,-4861){\line( 1, 0){1200}}
\put(3601,-2461){\line( 1, 0){1200}}
\put(3901,-2761){\line( 1, 0){900}}
\put(4801,-2761){\line( 0, 1){  0}}
\put(5701,-4561){\vector( 0, 1){900}}
\put(3901,-2761){\vector( 0,-1){900}}
\put(3901,-3661){\line( 0,-1){900}}
\put(3901,-4561){\vector( 1, 0){900}}
\put(4801,-4561){\line( 1, 0){900}}
\put(5701,-4561){\line( 0, 1){  0}}
\put(5701,-3661){\line( 0, 1){1200}}
\put(6001,-2761){\vector(-1, 0){1200}}
\put(6001,-2761){\line( 0,-1){900}}
\put(5701,-2461){\vector(-1, 0){900}}
\end{picture}
 \>  \\ \\ \\  
 \> 2 \> \setlength{\unitlength}{0.00083300in}%
\setlength{\unitlength}{0.00020in}%
\begingroup\makeatletter\ifx\SetFigFont\undefined
\def\x#1#2#3#4#5#6#7\relax{\def\x{#1#2#3#4#5#6}}%
\expandafter\x\fmtname xxxxxx\relax \def\y{splain}%
\ifx\x\y   
\gdef\SetFigFont#1#2#3{%
  \ifnum #1<17\tiny\else \ifnum #1<20\small\else
  \ifnum #1<24\normalsize\else \ifnum #1<29\large\else
  \ifnum #1<34\Large\else \ifnum #1<41\LARGE\else
     \huge\fi\fi\fi\fi\fi\fi
  \csname #3\endcsname}%
\else
\gdef\SetFigFont#1#2#3{\begingroup
  \count@#1\relax \ifnum 25<\count@\count@25\fi
  \def\x{\endgroup\@setsize\SetFigFont{#2pt}}%
  \expandafter\x
    \csname \romannumeral\the\count@ pt\expandafter\endcsname
    \csname @\romannumeral\the\count@ pt\endcsname
  \csname #3\endcsname}%
\fi
\fi\endgroup
\begin{picture}(1144,1444)(3579,-4883)
\thicklines
\put(3601,-4861){\vector( 1, 0){1200}}
\put(6001,-4861){\vector( 0, 1){1200}}
\put(3601,-4861){\line( 0, 1){1200}}
\put(4801,-4861){\line( 1, 0){1200}}
\put(6001,-2461){\vector(-1, 0){1200}}
\put(3601,-2461){\line( 1, 0){1200}}
\put(3601,-2461){\vector( 0,-1){1200}}
\put(6001,-3661){\line( 0, 1){1200}}
\put(6301,-2461){\vector( 0,-1){1200}}
\put(6301,-4861){\line( 0, 1){1200}}
\put(6301,-4861){\vector( 1, 0){1200}}
\put(7501,-4861){\line( 1, 0){1200}}
\put(6301,-2461){\line( 1, 0){1200}}
\put(8701,-2461){\vector(-1, 0){1200}}
\put(8701,-3661){\line( 0, 1){1200}}
\put(8701,-4861){\vector( 0, 1){1200}}
\end{picture}
 \>\hskip 3cm \setlength{\unitlength}{0.00083300in}%
\setlength{\unitlength}{0.000200in}%
\begingroup\makeatletter\ifx\SetFigFont\undefined
\def\x#1#2#3#4#5#6#7\relax{\def\x{#1#2#3#4#5#6}}%
\expandafter\x\fmtname xxxxxx\relax \def\y{splain}%
\ifx\x\y   
\gdef\SetFigFont#1#2#3{%
  \ifnum #1<17\tiny\else \ifnum #1<20\small\else
  \ifnum #1<24\normalsize\else \ifnum #1<29\large\else
  \ifnum #1<34\Large\else \ifnum #1<41\LARGE\else
     \huge\fi\fi\fi\fi\fi\fi
  \csname #3\endcsname}%
\else
\gdef\SetFigFont#1#2#3{\begingroup
  \count@#1\relax \ifnum 25<\count@\count@25\fi
  \def\x{\endgroup\@setsize\SetFigFont{#2pt}}%
  \expandafter\x
    \csname \romannumeral\the\count@ pt\expandafter\endcsname
    \csname @\romannumeral\the\count@ pt\endcsname
  \csname #3\endcsname}%
\fi
\fi\endgroup
\begin{picture}(4844,2444)(3579,-4883)
\thicklines
\put(3601,-4861){\vector( 1, 0){1200}}
\put(3601,-4861){\line( 0, 1){1200}}
\put(4801,-4861){\line( 1, 0){1200}}
\put(6001,-2461){\vector(-1, 0){1200}}
\put(3601,-2461){\line( 1, 0){1200}}
\put(3601,-2461){\vector( 0,-1){1200}}
\put(6001,-4861){\vector( 1, 0){1200}}
\put(7201,-4861){\line( 1, 0){1200}}
\put(6001,-2461){\line( 1, 0){1200}}
\put(8401,-2461){\vector(-1, 0){1200}}
\put(8401,-3661){\line( 0, 1){1200}}
\put(8401,-4861){\vector( 0, 1){1200}}
\end{picture}
 \>  \\ \\ \\
3\> 1 \> \setlength{\unitlength}{0.00083300in}%
\setlength{\unitlength}{0.000200in}%
\begingroup\makeatletter\ifx\SetFigFont\undefined
\def\x#1#2#3#4#5#6#7\relax{\def\x{#1#2#3#4#5#6}}%
\expandafter\x\fmtname xxxxxx\relax \def\y{splain}%
\ifx\x\y   
\gdef\SetFigFont#1#2#3{%
  \ifnum #1<17\tiny\else \ifnum #1<20\small\else
  \ifnum #1<24\normalsize\else \ifnum #1<29\large\else
  \ifnum #1<34\Large\else \ifnum #1<41\LARGE\else
     \huge\fi\fi\fi\fi\fi\fi
  \csname #3\endcsname}%
\else
\gdef\SetFigFont#1#2#3{\begingroup
  \count@#1\relax \ifnum 25<\count@\count@25\fi
  \def\x{\endgroup\@setsize\SetFigFont{#2pt}}%
  \expandafter\x
    \csname \romannumeral\the\count@ pt\expandafter\endcsname
    \csname @\romannumeral\the\count@ pt\endcsname
  \csname #3\endcsname}%
\fi
\fi\endgroup
\begin{picture}(2444,2444)(3579,-4883)
\thicklines
\put(3601,-4861){\vector( 1, 0){1200}}
\put(6001,-4861){\vector( 0, 1){1200}}
\put(6001,-3661){\line( 0, 1){1200}}
\put(3601,-2461){\vector( 0,-1){1200}}
\put(3601,-4861){\line( 0, 1){1200}}
\put(6001,-2461){\vector(-1, 0){1200}}
\put(4801,-4861){\line( 1, 0){1200}}
\put(3601,-2461){\line( 1, 0){1200}}
\put(3901,-2761){\line( 1, 0){900}}
\put(4801,-2761){\line( 0, 1){  0}}
\put(5701,-2761){\vector(-1, 0){900}}
\put(5701,-2761){\line( 0,-1){900}}
\put(5701,-4561){\vector( 0, 1){900}}
\put(3901,-2761){\vector( 0,-1){900}}
\put(3901,-3661){\line( 0,-1){900}}
\put(3901,-4561){\vector( 1, 0){900}}
\put(4801,-4561){\line( 1, 0){900}}
\put(5701,-4561){\line( 0, 1){  0}}
\put(5851,-2611){\vector(-1, 0){1050}}
\put(5851,-3661){\line( 0, 1){1050}}
\put(5851,-4711){\vector( 0, 1){1050}}
\put(3751,-2611){\line( 1, 0){1050}}
\put(3751,-2611){\vector( 0,-1){1050}}
\put(3751,-3511){\line( 0,-1){1200}}
\put(3751,-4711){\vector( 1, 0){1050}}
\put(4801,-4711){\line( 1, 0){1050}}
\end{picture}
 \>\hskip 3cm \setlength{\unitlength}{0.00083300in}%
\setlength{\unitlength}{0.000200in}%
\begingroup\makeatletter\ifx\SetFigFont\undefined
\def\x#1#2#3#4#5#6#7\relax{\def\x{#1#2#3#4#5#6}}%
\expandafter\x\fmtname xxxxxx\relax \def\y{splain}%
\ifx\x\y   
\gdef\SetFigFont#1#2#3{%
  \ifnum #1<17\tiny\else \ifnum #1<20\small\else
  \ifnum #1<24\normalsize\else \ifnum #1<29\large\else
  \ifnum #1<34\Large\else \ifnum #1<41\LARGE\else
     \huge\fi\fi\fi\fi\fi\fi
  \csname #3\endcsname}%
\else
\gdef\SetFigFont#1#2#3{\begingroup
  \count@#1\relax \ifnum 25<\count@\count@25\fi
  \def\x{\endgroup\@setsize\SetFigFont{#2pt}}%
  \expandafter\x
    \csname \romannumeral\the\count@ pt\expandafter\endcsname
    \csname @\romannumeral\the\count@ pt\endcsname
  \csname #3\endcsname}%
\fi
\fi\endgroup
\begin{picture}(2444,2444)(3579,-4883)
\thicklines
\put(3601,-4861){\vector( 1, 0){1200}}
\put(6001,-4861){\vector( 0, 1){1200}}
\put(6001,-3661){\line( 0, 1){1200}}
\put(3601,-2461){\vector( 0,-1){1200}}
\put(3601,-4861){\line( 0, 1){1200}}
\put(6001,-2461){\vector(-1, 0){1200}}
\put(4801,-4861){\line( 1, 0){1200}}
\put(3601,-2461){\line( 1, 0){1200}}
\put(3901,-2761){\line( 1, 0){900}}
\put(4801,-2761){\line( 0, 1){  0}}
\put(5701,-2761){\vector(-1, 0){900}}
\put(5701,-2761){\line( 0,-1){900}}
\put(3901,-2761){\vector( 0,-1){900}}
\put(3901,-3661){\line( 0,-1){900}}
\put(3901,-4561){\vector( 1, 0){900}}
\put(5851,-2611){\vector(-1, 0){1050}}
\put(5851,-3661){\line( 0, 1){1050}}
\put(3751,-2611){\line( 1, 0){1050}}
\put(3751,-2611){\vector( 0,-1){1050}}
\put(3751,-3511){\line( 0,-1){1200}}
\put(3751,-4711){\vector( 1, 0){1050}}
\put(4801,-4561){\line( 1, 0){1050}}
\put(5701,-4711){\vector( 0, 1){1050}}
\put(5851,-4561){\vector( 0, 1){900}}
\put(4801,-4711){\line( 1, 0){900}}
\put(5701,-4711){\line( 0, 1){  0}}
\end{picture}
 \>\hskip
3cm \setlength{\unitlength}{0.00083300in}%
\setlength{\unitlength}{0.000200in}%
\begingroup\makeatletter\ifx\SetFigFont\undefined
\def\x#1#2#3#4#5#6#7\relax{\def\x{#1#2#3#4#5#6}}%
\expandafter\x\fmtname xxxxxx\relax \def\y{splain}%
\ifx\x\y   
\gdef\SetFigFont#1#2#3{%
  \ifnum #1<17\tiny\else \ifnum #1<20\small\else
  \ifnum #1<24\normalsize\else \ifnum #1<29\large\else
  \ifnum #1<34\Large\else \ifnum #1<41\LARGE\else
     \huge\fi\fi\fi\fi\fi\fi
  \csname #3\endcsname}%
\else
\gdef\SetFigFont#1#2#3{\begingroup
  \count@#1\relax \ifnum 25<\count@\count@25\fi
  \def\x{\endgroup\@setsize\SetFigFont{#2pt}}%
  \expandafter\x
    \csname \romannumeral\the\count@ pt\expandafter\endcsname
    \csname @\romannumeral\the\count@ pt\endcsname
  \csname #3\endcsname}%
\fi
\fi\endgroup
\begin{picture}(2444,2444)(3579,-4883)
\thicklines
\put(3601,-4861){\vector( 1, 0){1200}}
\put(6001,-3661){\line( 0, 1){1200}}
\put(3601,-2461){\vector( 0,-1){1200}}
\put(3601,-4861){\line( 0, 1){1200}}
\put(6001,-2461){\vector(-1, 0){1200}}
\put(3601,-2461){\line( 1, 0){1200}}
\put(3901,-2761){\line( 1, 0){900}}
\put(4801,-2761){\line( 0, 1){  0}}
\put(5701,-2761){\vector(-1, 0){900}}
\put(5701,-2761){\line( 0,-1){900}}
\put(3901,-2761){\vector( 0,-1){900}}
\put(3901,-3661){\line( 0,-1){900}}
\put(3901,-4561){\vector( 1, 0){900}}
\put(5851,-2611){\vector(-1, 0){1050}}
\put(5851,-3661){\line( 0, 1){1050}}
\put(3751,-2611){\line( 1, 0){1050}}
\put(3751,-2611){\vector( 0,-1){1050}}
\put(3751,-3511){\line( 0,-1){1200}}
\put(3751,-4711){\vector( 1, 0){1050}}
\put(4801,-4561){\line( 1, 0){1200}}
\put(4801,-4861){\line( 1, 0){1050}}
\put(5851,-4861){\vector( 0, 1){1200}}
\put(5701,-4711){\vector( 0, 1){1050}}
\put(4801,-4711){\line( 1, 0){900}}
\put(5701,-4711){\line( 0, 1){  0}}
\put(6001,-4561){\vector( 0, 1){900}}
\end{picture}
  \\ \\ \\ \\ 
 \> 2 \> \setlength{\unitlength}{0.00083300in}%
\setlength{\unitlength}{0.000200in}%
\begingroup\makeatletter\ifx\SetFigFont\undefined
\def\x#1#2#3#4#5#6#7\relax{\def\x{#1#2#3#4#5#6}}%
\expandafter\x\fmtname xxxxxx\relax \def\y{splain}%
\ifx\x\y   
\gdef\SetFigFont#1#2#3{%
  \ifnum #1<17\tiny\else \ifnum #1<20\small\else
  \ifnum #1<24\normalsize\else \ifnum #1<29\large\else
  \ifnum #1<34\Large\else \ifnum #1<41\LARGE\else
     \huge\fi\fi\fi\fi\fi\fi
  \csname #3\endcsname}%
\else
\gdef\SetFigFont#1#2#3{\begingroup
  \count@#1\relax \ifnum 25<\count@\count@25\fi
  \def\x{\endgroup\@setsize\SetFigFont{#2pt}}%
  \expandafter\x
    \csname \romannumeral\the\count@ pt\expandafter\endcsname
    \csname @\romannumeral\the\count@ pt\endcsname
  \csname #3\endcsname}%
\fi
\fi\endgroup
\begin{picture}(4844,2444)(3579,-4883)
\thicklines
\put(3601,-4861){\vector( 1, 0){1200}}
\put(6001,-4861){\vector( 0, 1){1200}}
\put(6001,-3661){\line( 0, 1){1200}}
\put(3601,-2461){\vector( 0,-1){1200}}
\put(3601,-4861){\line( 0, 1){1200}}
\put(6001,-2461){\vector(-1, 0){1200}}
\put(4801,-4861){\line( 1, 0){1200}}
\put(3601,-2461){\line( 1, 0){1200}}
\put(3901,-2761){\line( 1, 0){900}}
\put(4801,-2761){\line( 0, 1){  0}}
\put(5701,-2761){\vector(-1, 0){900}}
\put(5701,-2761){\line( 0,-1){900}}
\put(5701,-4561){\vector( 0, 1){900}}
\put(3901,-2761){\vector( 0,-1){900}}
\put(3901,-3661){\line( 0,-1){900}}
\put(3901,-4561){\vector( 1, 0){900}}
\put(4801,-4561){\line( 1, 0){900}}
\put(5701,-4561){\line( 0, 1){  0}}
\put(6301,-2611){\vector( 0,-1){1050}}
\put(6301,-3511){\line( 0,-1){1200}}
\put(8401,-3661){\line( 0, 1){1050}}
\put(8401,-4711){\vector( 0, 1){1050}}
\put(7351,-4711){\line( 1, 0){1050}}
\put(8401,-2611){\vector(-1, 0){1050}}
\put(6301,-2611){\line( 1, 0){1050}}
\put(6301,-4711){\vector( 1, 0){1050}}
\end{picture}
 \>\hskip 3cm \setlength{\unitlength}{0.00083300in}%
\setlength{\unitlength}{0.0002000in}%
\begingroup\makeatletter\ifx\SetFigFont\undefined
\def\x#1#2#3#4#5#6#7\relax{\def\x{#1#2#3#4#5#6}}%
\expandafter\x\fmtname xxxxxx\relax \def\y{splain}%
\ifx\x\y   
\gdef\SetFigFont#1#2#3{%
  \ifnum #1<17\tiny\else \ifnum #1<20\small\else
  \ifnum #1<24\normalsize\else \ifnum #1<29\large\else
  \ifnum #1<34\Large\else \ifnum #1<41\LARGE\else
     \huge\fi\fi\fi\fi\fi\fi
  \csname #3\endcsname}%
\else
\gdef\SetFigFont#1#2#3{\begingroup
  \count@#1\relax \ifnum 25<\count@\count@25\fi
  \def\x{\endgroup\@setsize\SetFigFont{#2pt}}%
  \expandafter\x
    \csname \romannumeral\the\count@ pt\expandafter\endcsname
    \csname @\romannumeral\the\count@ pt\endcsname
  \csname #3\endcsname}%
\fi
\fi\endgroup
\begin{picture}(4844,2444)(3579,-4883)
\thicklines
\put(3601,-4861){\vector( 1, 0){1200}}
\put(6001,-4861){\vector( 0, 1){1200}}
\put(6001,-3661){\line( 0, 1){1200}}
\put(3601,-4861){\line( 0, 1){1200}}
\put(6001,-2461){\vector(-1, 0){1200}}
\put(4801,-4861){\line( 1, 0){1200}}
\put(5701,-2761){\vector(-1, 0){900}}
\put(5701,-2761){\line( 0,-1){900}}
\put(5701,-4561){\vector( 0, 1){900}}
\put(3901,-3661){\line( 0,-1){900}}
\put(3901,-4561){\vector( 1, 0){900}}
\put(4801,-4561){\line( 1, 0){900}}
\put(5701,-4561){\line( 0, 1){  0}}
\put(6301,-2611){\vector( 0,-1){1050}}
\put(6301,-3511){\line( 0,-1){1200}}
\put(8401,-3661){\line( 0, 1){1050}}
\put(8401,-4711){\vector( 0, 1){1050}}
\put(7351,-4711){\line( 1, 0){1050}}
\put(8401,-2611){\vector(-1, 0){1050}}
\put(6301,-2611){\line( 1, 0){1050}}
\put(6301,-4711){\vector( 1, 0){1050}}
\put(3601,-2761){\line( 1, 0){1200}}
\put(3901,-2461){\vector( 0,-1){1200}}
\put(3901,-2461){\line( 1, 0){900}}
\put(4801,-2461){\line( 0, 1){  0}}
\put(3601,-2761){\vector( 0,-1){900}}
\end{picture}
 \>\hskip
3cm \setlength{\unitlength}{0.00083300in}%
\setlength{\unitlength}{0.000200in}%
\begingroup\makeatletter\ifx\SetFigFont\undefined
\def\x#1#2#3#4#5#6#7\relax{\def\x{#1#2#3#4#5#6}}%
\expandafter\x\fmtname xxxxxx\relax \def\y{splain}%
\ifx\x\y   
\gdef\SetFigFont#1#2#3{%
  \ifnum #1<17\tiny\else \ifnum #1<20\small\else
  \ifnum #1<24\normalsize\else \ifnum #1<29\large\else
  \ifnum #1<34\Large\else \ifnum #1<41\LARGE\else
     \huge\fi\fi\fi\fi\fi\fi
  \csname #3\endcsname}%
\else
\gdef\SetFigFont#1#2#3{\begingroup
  \count@#1\relax \ifnum 25<\count@\count@25\fi
  \def\x{\endgroup\@setsize\SetFigFont{#2pt}}%
  \expandafter\x
    \csname \romannumeral\the\count@ pt\expandafter\endcsname
    \csname @\romannumeral\the\count@ pt\endcsname
  \csname #3\endcsname}%
\fi
\fi\endgroup
\begin{picture}(4844,2444)(3579,-4883)
\thicklines
\put(3601,-4861){\vector( 1, 0){1200}}
\put(3601,-2461){\vector( 0,-1){1200}}
\put(3601,-4861){\line( 0, 1){1200}}
\put(6001,-2461){\vector(-1, 0){1200}}
\put(4801,-4861){\line( 1, 0){1200}}
\put(3601,-2461){\line( 1, 0){1200}}
\put(3901,-2761){\line( 1, 0){900}}
\put(4801,-2761){\line( 0, 1){  0}}
\put(5701,-2761){\vector(-1, 0){900}}
\put(5701,-2761){\line( 0,-1){900}}
\put(5701,-4561){\vector( 0, 1){900}}
\put(3901,-2761){\vector( 0,-1){900}}
\put(3901,-3661){\line( 0,-1){900}}
\put(3901,-4561){\vector( 1, 0){900}}
\put(4801,-4561){\line( 1, 0){900}}
\put(5701,-4561){\line( 0, 1){  0}}
\put(6001,-4861){\vector( 1, 0){1200}}
\put(7201,-4861){\line( 1, 0){1200}}
\put(6001,-2461){\line( 1, 0){1200}}
\put(8401,-2461){\vector(-1, 0){1200}}
\put(8401,-3661){\line( 0, 1){1200}}
\put(8401,-4861){\vector( 0, 1){1200}}
\end{picture}
 \> \hskip 3cm \setlength{\unitlength}{0.00083300in}%
\setlength{\unitlength}{0.000200in}%
\begingroup\makeatletter\ifx\SetFigFont\undefined
\def\x#1#2#3#4#5#6#7\relax{\def\x{#1#2#3#4#5#6}}%
\expandafter\x\fmtname xxxxxx\relax \def\y{splain}%
\ifx\x\y   
\gdef\SetFigFont#1#2#3{%
  \ifnum #1<17\tiny\else \ifnum #1<20\small\else
  \ifnum #1<24\normalsize\else \ifnum #1<29\large\else
  \ifnum #1<34\Large\else \ifnum #1<41\LARGE\else
     \huge\fi\fi\fi\fi\fi\fi
  \csname #3\endcsname}%
\else
\gdef\SetFigFont#1#2#3{\begingroup
  \count@#1\relax \ifnum 25<\count@\count@25\fi
  \def\x{\endgroup\@setsize\SetFigFont{#2pt}}%
  \expandafter\x
    \csname \romannumeral\the\count@ pt\expandafter\endcsname
    \csname @\romannumeral\the\count@ pt\endcsname
  \csname #3\endcsname}%
\fi
\fi\endgroup
\begin{picture}(4844,2444)(3579,-4883)
\thicklines
\put(3601,-4861){\vector( 1, 0){1200}}
\put(3601,-4861){\line( 0, 1){1200}}
\put(6001,-2461){\vector(-1, 0){1200}}
\put(4801,-4861){\line( 1, 0){1200}}
\put(5701,-2761){\vector(-1, 0){900}}
\put(5701,-2761){\line( 0,-1){900}}
\put(5701,-4561){\vector( 0, 1){900}}
\put(3901,-3661){\line( 0,-1){900}}
\put(3901,-4561){\vector( 1, 0){900}}
\put(4801,-4561){\line( 1, 0){900}}
\put(5701,-4561){\line( 0, 1){  0}}
\put(6001,-4861){\vector( 1, 0){1200}}
\put(7201,-4861){\line( 1, 0){1200}}
\put(6001,-2461){\line( 1, 0){1200}}
\put(8401,-2461){\vector(-1, 0){1200}}
\put(8401,-3661){\line( 0, 1){1200}}
\put(8401,-4861){\vector( 0, 1){1200}}
\put(3901,-2461){\vector( 0,-1){1200}}
\put(3601,-2761){\line( 1, 0){1200}}
\put(3901,-2461){\line( 1, 0){900}}
\put(4801,-2461){\line( 0, 1){  0}}
\put(3601,-2761){\vector( 0,-1){900}}
\end{picture}
 \> \\   \\ \\ 
 \> 3 \> \setlength{\unitlength}{0.00083300in}%
\setlength{\unitlength}{0.000150in}%
\begingroup\makeatletter\ifx\SetFigFont\undefined
\def\x#1#2#3#4#5#6#7\relax{\def\x{#1#2#3#4#5#6}}%
\expandafter\x\fmtname xxxxxx\relax \def\y{splain}%
\ifx\x\y   
\gdef\SetFigFont#1#2#3{%
  \ifnum #1<17\tiny\else \ifnum #1<20\small\else
  \ifnum #1<24\normalsize\else \ifnum #1<29\large\else
  \ifnum #1<34\Large\else \ifnum #1<41\LARGE\else
     \huge\fi\fi\fi\fi\fi\fi
  \csname #3\endcsname}%
\else
\gdef\SetFigFont#1#2#3{\begingroup
  \count@#1\relax \ifnum 25<\count@\count@25\fi
  \def\x{\endgroup\@setsize\SetFigFont{#2pt}}%
  \expandafter\x
    \csname \romannumeral\the\count@ pt\expandafter\endcsname
    \csname @\romannumeral\the\count@ pt\endcsname
  \csname #3\endcsname}%
\fi
\fi\endgroup
\begin{picture}(7844,2444)(3579,-4883)
\thicklines
\put(3601,-4861){\vector( 1, 0){1200}}
\put(6001,-4861){\vector( 0, 1){1200}}
\put(3601,-4861){\line( 0, 1){1200}}
\put(4801,-4861){\line( 1, 0){1200}}
\put(6001,-2461){\vector(-1, 0){1200}}
\put(3601,-2461){\line( 1, 0){1200}}
\put(3601,-2461){\vector( 0,-1){1200}}
\put(6001,-3661){\line( 0, 1){1200}}
\put(6301,-2461){\vector( 0,-1){1200}}
\put(6301,-4861){\line( 0, 1){1200}}
\put(6301,-4861){\vector( 1, 0){1200}}
\put(7501,-4861){\line( 1, 0){1200}}
\put(6301,-2461){\line( 1, 0){1200}}
\put(8701,-2461){\vector(-1, 0){1200}}
\put(8701,-3661){\line( 0, 1){1200}}
\put(8701,-4861){\vector( 0, 1){1200}}
\put(9001,-2461){\vector( 0,-1){1200}}
\put(9001,-4861){\line( 0, 1){1200}}
\put(9001,-4861){\vector( 1, 0){1200}}
\put(10201,-4861){\line( 1, 0){1200}}
\put(9001,-2461){\line( 1, 0){1200}}
\put(11401,-2461){\vector(-1, 0){1200}}
\put(11401,-3661){\line( 0, 1){1200}}
\put(11401,-4861){\vector( 0, 1){1200}}
\end{picture}
 \>\hskip 3cm \setlength{\unitlength}{0.00083300in}%
\setlength{\unitlength}{0.000150in}%
\begingroup\makeatletter\ifx\SetFigFont\undefined
\def\x#1#2#3#4#5#6#7\relax{\def\x{#1#2#3#4#5#6}}%
\expandafter\x\fmtname xxxxxx\relax \def\y{splain}%
\ifx\x\y   
\gdef\SetFigFont#1#2#3{%
  \ifnum #1<17\tiny\else \ifnum #1<20\small\else
  \ifnum #1<24\normalsize\else \ifnum #1<29\large\else
  \ifnum #1<34\Large\else \ifnum #1<41\LARGE\else
     \huge\fi\fi\fi\fi\fi\fi
  \csname #3\endcsname}%
\else
\gdef\SetFigFont#1#2#3{\begingroup
  \count@#1\relax \ifnum 25<\count@\count@25\fi
  \def\x{\endgroup\@setsize\SetFigFont{#2pt}}%
  \expandafter\x
    \csname \romannumeral\the\count@ pt\expandafter\endcsname
    \csname @\romannumeral\the\count@ pt\endcsname
  \csname #3\endcsname}%
\fi
\fi\endgroup
\begin{picture}(7544,2444)(3879,-4883)
\thicklines
\put(6301,-4861){\vector( 1, 0){1200}}
\put(7501,-4861){\line( 1, 0){1200}}
\put(6301,-2461){\line( 1, 0){1200}}
\put(8701,-2461){\vector(-1, 0){1200}}
\put(8701,-3661){\line( 0, 1){1200}}
\put(8701,-4861){\vector( 0, 1){1200}}
\put(9001,-2461){\vector( 0,-1){1200}}
\put(9001,-4861){\line( 0, 1){1200}}
\put(9001,-4861){\vector( 1, 0){1200}}
\put(10201,-4861){\line( 1, 0){1200}}
\put(9001,-2461){\line( 1, 0){1200}}
\put(11401,-2461){\vector(-1, 0){1200}}
\put(11401,-3661){\line( 0, 1){1200}}
\put(11401,-4861){\vector( 0, 1){1200}}
\put(6301,-2461){\vector(-1, 0){1200}}
\put(3901,-2461){\line( 1, 0){1200}}
\put(3901,-2461){\vector( 0,-1){1200}}
\put(3901,-4861){\line( 0, 1){1200}}
\put(5101,-4861){\line( 1, 0){1200}}
\put(3901,-4861){\vector( 1, 0){1200}}
\end{picture}
 \>\hskip
3.2cm \setlength{\unitlength}{0.00083300in}%
\setlength{\unitlength}{0.000150in}%
\begingroup\makeatletter\ifx\SetFigFont\undefined
\def\x#1#2#3#4#5#6#7\relax{\def\x{#1#2#3#4#5#6}}%
\expandafter\x\fmtname xxxxxx\relax \def\y{splain}%
\ifx\x\y   
\gdef\SetFigFont#1#2#3{%
  \ifnum #1<17\tiny\else \ifnum #1<20\small\else
  \ifnum #1<24\normalsize\else \ifnum #1<29\large\else
  \ifnum #1<34\Large\else \ifnum #1<41\LARGE\else
     \huge\fi\fi\fi\fi\fi\fi
  \csname #3\endcsname}%
\else
\gdef\SetFigFont#1#2#3{\begingroup
  \count@#1\relax \ifnum 25<\count@\count@25\fi
  \def\x{\endgroup\@setsize\SetFigFont{#2pt}}%
  \expandafter\x
    \csname \romannumeral\the\count@ pt\expandafter\endcsname
    \csname @\romannumeral\the\count@ pt\endcsname
  \csname #3\endcsname}%
\fi
\fi\endgroup
\begin{picture}(7544,2444)(3579,-4883)
\thicklines
\put(3601,-4861){\vector( 1, 0){1200}}
\put(6001,-4861){\vector( 0, 1){1200}}
\put(3601,-4861){\line( 0, 1){1200}}
\put(4801,-4861){\line( 1, 0){1200}}
\put(6001,-2461){\vector(-1, 0){1200}}
\put(3601,-2461){\line( 1, 0){1200}}
\put(3601,-2461){\vector( 0,-1){1200}}
\put(6001,-3661){\line( 0, 1){1200}}
\put(6301,-2461){\vector( 0,-1){1200}}
\put(6301,-4861){\line( 0, 1){1200}}
\put(6301,-4861){\vector( 1, 0){1200}}
\put(7501,-4861){\line( 1, 0){1200}}
\put(6301,-2461){\line( 1, 0){1200}}
\put(8701,-2461){\vector(-1, 0){1200}}
\put(8701,-2461){\line( 1, 0){1200}}
\put(11101,-2461){\vector(-1, 0){1200}}
\put(11101,-3661){\line( 0, 1){1200}}
\put(11101,-4861){\vector( 0, 1){1200}}
\put(8701,-4861){\vector( 1, 0){1200}}
\put(9901,-4861){\line( 1, 0){1200}}
\end{picture}
 \> \hskip 3.2cm \setlength{\unitlength}{0.00083300in}%
\setlength{\unitlength}{0.000150in}%
\begingroup\makeatletter\ifx\SetFigFont\undefined
\def\x#1#2#3#4#5#6#7\relax{\def\x{#1#2#3#4#5#6}}%
\expandafter\x\fmtname xxxxxx\relax \def\y{splain}%
\ifx\x\y   
\gdef\SetFigFont#1#2#3{%
  \ifnum #1<17\tiny\else \ifnum #1<20\small\else
  \ifnum #1<24\normalsize\else \ifnum #1<29\large\else
  \ifnum #1<34\Large\else \ifnum #1<41\LARGE\else
     \huge\fi\fi\fi\fi\fi\fi
  \csname #3\endcsname}%
\else
\gdef\SetFigFont#1#2#3{\begingroup
  \count@#1\relax \ifnum 25<\count@\count@25\fi
  \def\x{\endgroup\@setsize\SetFigFont{#2pt}}%
  \expandafter\x
    \csname \romannumeral\the\count@ pt\expandafter\endcsname
    \csname @\romannumeral\the\count@ pt\endcsname
  \csname #3\endcsname}%
\fi
\fi\endgroup
\begin{picture}(7244,2444)(3879,-4883)
\thicklines
\put(6301,-4861){\vector( 1, 0){1200}}
\put(7501,-4861){\line( 1, 0){1200}}
\put(6301,-2461){\line( 1, 0){1200}}
\put(8701,-2461){\vector(-1, 0){1200}}
\put(6301,-2461){\vector(-1, 0){1200}}
\put(3901,-2461){\line( 1, 0){1200}}
\put(3901,-2461){\vector( 0,-1){1200}}
\put(3901,-4861){\line( 0, 1){1200}}
\put(5101,-4861){\line( 1, 0){1200}}
\put(3901,-4861){\vector( 1, 0){1200}}
\put(8701,-2461){\line( 1, 0){1200}}
\put(11101,-2461){\vector(-1, 0){1200}}
\put(8701,-4861){\vector( 1, 0){1200}}
\put(9901,-4861){\line( 1, 0){1200}}
\put(11101,-4861){\vector( 0, 1){1200}}
\put(11101,-3661){\line( 0, 1){1200}}
\end{picture}
 \> \\ \\ \\
 \> 4 \> \setlength{\unitlength}{0.00083300in}%
\setlength{\unitlength}{0.000200in}%
\begingroup\makeatletter\ifx\SetFigFont\undefined
\def\x#1#2#3#4#5#6#7\relax{\def\x{#1#2#3#4#5#6}}%
\expandafter\x\fmtname xxxxxx\relax \def\y{splain}%
\ifx\x\y   
\gdef\SetFigFont#1#2#3{%
  \ifnum #1<17\tiny\else \ifnum #1<20\small\else
  \ifnum #1<24\normalsize\else \ifnum #1<29\large\else
  \ifnum #1<34\Large\else \ifnum #1<41\LARGE\else
     \huge\fi\fi\fi\fi\fi\fi
  \csname #3\endcsname}%
\else
\gdef\SetFigFont#1#2#3{\begingroup
  \count@#1\relax \ifnum 25<\count@\count@25\fi
  \def\x{\endgroup\@setsize\SetFigFont{#2pt}}%
  \expandafter\x
    \csname \romannumeral\the\count@ pt\expandafter\endcsname
    \csname @\romannumeral\the\count@ pt\endcsname
  \csname #3\endcsname}%
\fi
\fi\endgroup
\begin{picture}(5144,5144)(3579,-4883)
\thicklines
\put(3601,-4861){\vector( 1, 0){1200}}
\put(6001,-4861){\vector( 0, 1){1200}}
\put(3601,-4861){\line( 0, 1){1200}}
\put(4801,-4861){\line( 1, 0){1200}}
\put(6001,-2461){\vector(-1, 0){1200}}
\put(3601,-2461){\line( 1, 0){1200}}
\put(3601,-2461){\vector( 0,-1){1200}}
\put(6001,-3661){\line( 0, 1){1200}}
\put(6301,-2461){\vector( 0,-1){1200}}
\put(6301,-4861){\line( 0, 1){1200}}
\put(6301,-4861){\vector( 1, 0){1200}}
\put(7501,-4861){\line( 1, 0){1200}}
\put(6301,-2461){\line( 1, 0){1200}}
\put(8701,-2461){\vector(-1, 0){1200}}
\put(8701,-3661){\line( 0, 1){1200}}
\put(8701,-4861){\vector( 0, 1){1200}}
\put(3601,239){\vector( 0,-1){1200}}
\put(3601,-2161){\line( 0, 1){1200}}
\put(3601,-2161){\vector( 1, 0){1200}}
\put(4801,-2161){\line( 1, 0){1200}}
\put(6001,-2161){\vector( 0, 1){1200}}
\put(6001,-961){\line( 0, 1){1200}}
\put(3601,239){\line( 1, 0){1200}}
\put(6001,239){\vector(-1, 0){1200}}
\end{picture}
 \>\hskip 3cm \setlength{\unitlength}{0.00083300in}%
\setlength{\unitlength}{0.000200in}%
\begingroup\makeatletter\ifx\SetFigFont\undefined
\def\x#1#2#3#4#5#6#7\relax{\def\x{#1#2#3#4#5#6}}%
\expandafter\x\fmtname xxxxxx\relax \def\y{splain}%
\ifx\x\y   
\gdef\SetFigFont#1#2#3{%
  \ifnum #1<17\tiny\else \ifnum #1<20\small\else
  \ifnum #1<24\normalsize\else \ifnum #1<29\large\else
  \ifnum #1<34\Large\else \ifnum #1<41\LARGE\else
     \huge\fi\fi\fi\fi\fi\fi
  \csname #3\endcsname}%
\else
\gdef\SetFigFont#1#2#3{\begingroup
  \count@#1\relax \ifnum 25<\count@\count@25\fi
  \def\x{\endgroup\@setsize\SetFigFont{#2pt}}%
  \expandafter\x
    \csname \romannumeral\the\count@ pt\expandafter\endcsname
    \csname @\romannumeral\the\count@ pt\endcsname
  \csname #3\endcsname}%
\fi
\fi\endgroup
\begin{picture}(5144,5144)(3579,-4883)
\thicklines
\put(3601,-4861){\vector( 1, 0){1200}}
\put(3601,-4861){\line( 0, 1){1200}}
\put(4801,-4861){\line( 1, 0){1200}}
\put(6001,-2461){\vector(-1, 0){1200}}
\put(3601,-2461){\line( 1, 0){1200}}
\put(3601,-2461){\vector( 0,-1){1200}}
\put(6301,-4861){\vector( 1, 0){1200}}
\put(7501,-4861){\line( 1, 0){1200}}
\put(6301,-2461){\line( 1, 0){1200}}
\put(8701,-2461){\vector(-1, 0){1200}}
\put(8701,-3661){\line( 0, 1){1200}}
\put(8701,-4861){\vector( 0, 1){1200}}
\put(3601,239){\vector( 0,-1){1200}}
\put(3601,-2161){\line( 0, 1){1200}}
\put(3601,-2161){\vector( 1, 0){1200}}
\put(4801,-2161){\line( 1, 0){1200}}
\put(6001,-2161){\vector( 0, 1){1200}}
\put(6001,-961){\line( 0, 1){1200}}
\put(3601,239){\line( 1, 0){1200}}
\put(6001,239){\vector(-1, 0){1200}}
\put(6001,-2461){\line( 1, 0){300}}
\put(6001,-4861){\line( 1, 0){300}}
\end{picture}
 \>\hskip
2.7cm \setlength{\unitlength}{0.00083300in}%
\setlength{\unitlength}{0.000200in}%
\begingroup\makeatletter\ifx\SetFigFont\undefined
\def\x#1#2#3#4#5#6#7\relax{\def\x{#1#2#3#4#5#6}}%
\expandafter\x\fmtname xxxxxx\relax \def\y{splain}%
\ifx\x\y   
\gdef\SetFigFont#1#2#3{%
  \ifnum #1<17\tiny\else \ifnum #1<20\small\else
  \ifnum #1<24\normalsize\else \ifnum #1<29\large\else
  \ifnum #1<34\Large\else \ifnum #1<41\LARGE\else
     \huge\fi\fi\fi\fi\fi\fi
  \csname #3\endcsname}%
\else
\gdef\SetFigFont#1#2#3{\begingroup
  \count@#1\relax \ifnum 25<\count@\count@25\fi
  \def\x{\endgroup\@setsize\SetFigFont{#2pt}}%
  \expandafter\x
    \csname \romannumeral\the\count@ pt\expandafter\endcsname
    \csname @\romannumeral\the\count@ pt\endcsname
  \csname #3\endcsname}%
\fi
\fi\endgroup
\begin{picture}(5144,5144)(3579,-4883)
\thicklines
\put(3601,-4861){\vector( 1, 0){1200}}
\put(6001,-4861){\vector( 0, 1){1200}}
\put(3601,-4861){\line( 0, 1){1200}}
\put(4801,-4861){\line( 1, 0){1200}}
\put(3601,-2461){\vector( 0,-1){1200}}
\put(6001,-3661){\line( 0, 1){1200}}
\put(6301,-2461){\vector( 0,-1){1200}}
\put(6301,-4861){\line( 0, 1){1200}}
\put(6301,-4861){\vector( 1, 0){1200}}
\put(7501,-4861){\line( 1, 0){1200}}
\put(6301,-2461){\line( 1, 0){1200}}
\put(8701,-2461){\vector(-1, 0){1200}}
\put(8701,-3661){\line( 0, 1){1200}}
\put(8701,-4861){\vector( 0, 1){1200}}
\put(3601,239){\vector( 0,-1){1200}}
\put(3601,-2161){\line( 0, 1){1200}}
\put(6001,-2161){\vector( 0, 1){1200}}
\put(6001,-961){\line( 0, 1){1200}}
\put(3601,239){\line( 1, 0){1200}}
\put(6001,239){\vector(-1, 0){1200}}
\put(3601,-2161){\line( 0,-1){300}}
\put(6001,-2161){\line( 0,-1){300}}
\end{picture}
 \> \hskip 2.7cm \setlength{\unitlength}{0.00083300in}%
\setlength{\unitlength}{0.000200in}%
\begingroup\makeatletter\ifx\SetFigFont\undefined
\def\x#1#2#3#4#5#6#7\relax{\def\x{#1#2#3#4#5#6}}%
\expandafter\x\fmtname xxxxxx\relax \def\y{splain}%
\ifx\x\y   
\gdef\SetFigFont#1#2#3{%
  \ifnum #1<17\tiny\else \ifnum #1<20\small\else
  \ifnum #1<24\normalsize\else \ifnum #1<29\large\else
  \ifnum #1<34\Large\else \ifnum #1<41\LARGE\else
     \huge\fi\fi\fi\fi\fi\fi
  \csname #3\endcsname}%
\else
\gdef\SetFigFont#1#2#3{\begingroup
  \count@#1\relax \ifnum 25<\count@\count@25\fi
  \def\x{\endgroup\@setsize\SetFigFont{#2pt}}%
  \expandafter\x
    \csname \romannumeral\the\count@ pt\expandafter\endcsname
    \csname @\romannumeral\the\count@ pt\endcsname
  \csname #3\endcsname}%
\fi
\fi\endgroup
\begin{picture}(4844,4844)(3579,-4883)
\thicklines
\put(3601,-4861){\vector( 1, 0){1200}}
\put(3601,-4861){\line( 0, 1){1200}}
\put(4801,-4861){\line( 1, 0){1200}}
\put(3601,-2461){\vector( 0,-1){1200}}
\put(3601,-2461){\line( 0, 1){1200}}
\put(3601,-61){\vector( 0,-1){1200}}
\put(3601,-61){\line( 1, 0){1200}}
\put(6001,-61){\vector(-1, 0){1200}}
\put(6001,-1261){\line( 0, 1){1200}}
\put(6001,-2461){\vector( 0, 1){1200}}
\put(6001,-2461){\line( 1, 0){1200}}
\put(6001,-4861){\vector( 1, 0){1200}}
\put(7201,-4861){\line( 1, 0){1200}}
\put(8401,-4861){\vector( 0, 1){1200}}
\put(8401,-3661){\line( 0, 1){1200}}
\put(8401,-2461){\vector(-1, 0){1200}}
\end{picture}
 \>\hskip
2.7cm \setlength{\unitlength}{0.00083300in}%
\setlength{\unitlength}{0.000200in}%
\begingroup\makeatletter\ifx\SetFigFont\undefined
\def\x#1#2#3#4#5#6#7\relax{\def\x{#1#2#3#4#5#6}}%
\expandafter\x\fmtname xxxxxx\relax \def\y{splain}%
\ifx\x\y   
\gdef\SetFigFont#1#2#3{%
  \ifnum #1<17\tiny\else \ifnum #1<20\small\else
  \ifnum #1<24\normalsize\else \ifnum #1<29\large\else
  \ifnum #1<34\Large\else \ifnum #1<41\LARGE\else
     \huge\fi\fi\fi\fi\fi\fi
  \csname #3\endcsname}%
\else
\gdef\SetFigFont#1#2#3{\begingroup
  \count@#1\relax \ifnum 25<\count@\count@25\fi
  \def\x{\endgroup\@setsize\SetFigFont{#2pt}}%
  \expandafter\x
    \csname \romannumeral\the\count@ pt\expandafter\endcsname
    \csname @\romannumeral\the\count@ pt\endcsname
  \csname #3\endcsname}%
\fi
\fi\endgroup
\begin{picture}(4994,4994)(3579,-4883)
\thicklines
\put(3601,-4861){\vector( 1, 0){1200}}
\put(6001,-4861){\vector( 0, 1){1200}}
\put(3601,-4861){\line( 0, 1){1200}}
\put(4801,-4861){\line( 1, 0){1200}}
\put(6001,-2461){\vector(-1, 0){1200}}
\put(3601,-2461){\line( 1, 0){1200}}
\put(3601,-2461){\vector( 0,-1){1200}}
\put(6001,-3661){\line( 0, 1){1200}}
\put(3601,-2311){\vector( 1, 0){1200}}
\put(3601, 89){\line( 1, 0){1200}}
\put(3601, 89){\vector( 0,-1){1200}}
\put(3601,-2311){\line( 0, 1){1200}}
\put(6151,-4861){\line( 0, 1){1200}}
\put(6151,-4861){\vector( 1, 0){1200}}
\put(7351,-4861){\line( 1, 0){1200}}
\put(8551,-4861){\vector( 0, 1){1200}}
\put(4801,-2311){\line( 1, 0){1350}}
\put(6151,-2311){\vector( 0,-1){1350}}
\put(6301,-1111){\line( 0, 1){1200}}
\put(8551,-2161){\vector(-1, 0){1200}}
\put(7351,-2161){\line(-1, 0){1050}}
\put(6301,-2161){\vector( 0, 1){1050}}
\put(6301, 89){\vector(-1, 0){1500}}
\put(8551,-3661){\line( 0, 1){1500}}
\end{picture}
 \> \\ \\ 
\end{tabbing}

Figure 4.
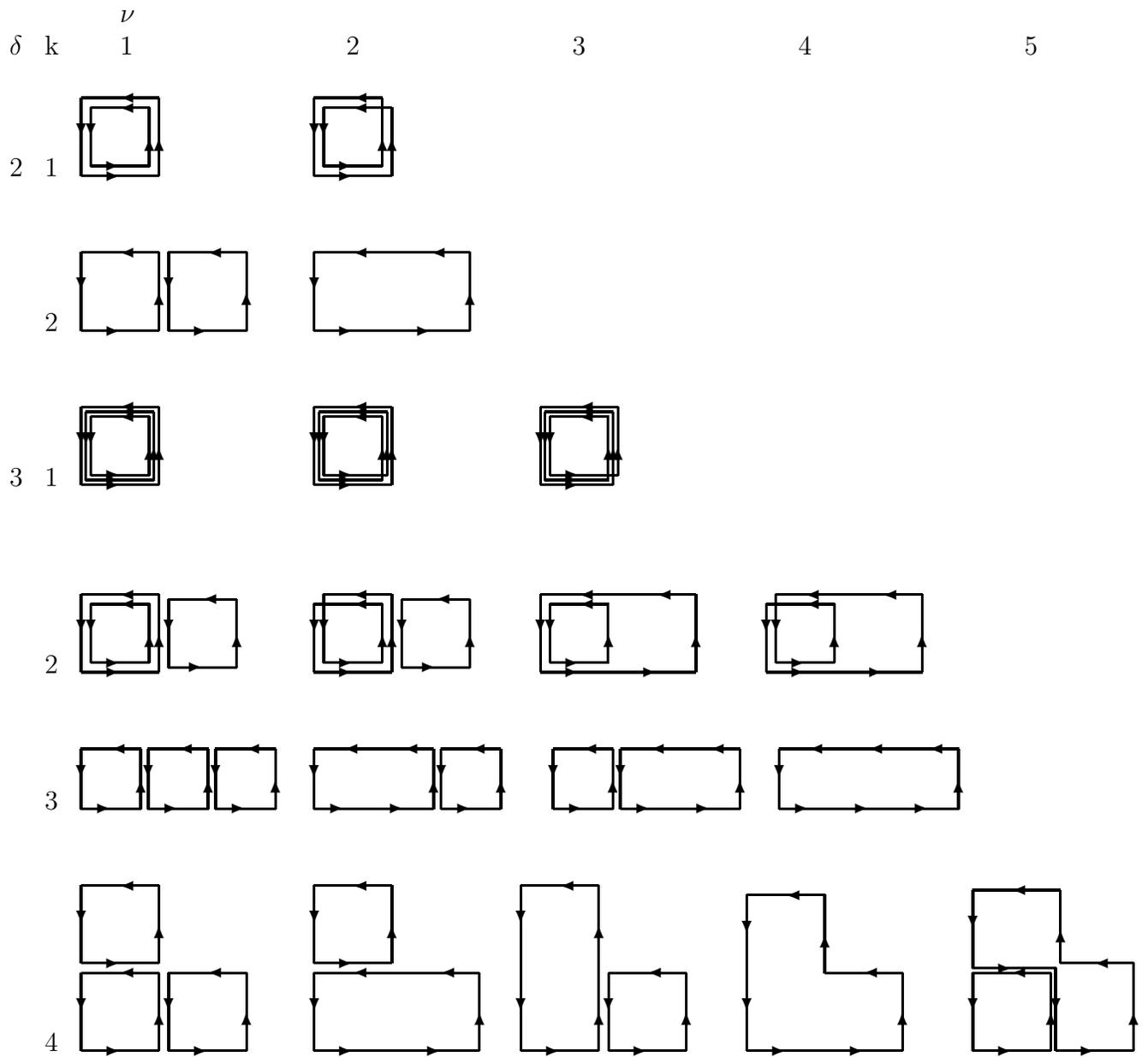
Loop structure of the functions $\Lambda^{\delta,k,\nu}$
(see equation  (54)) up to order $\delta = 3$.

\end{document}